\PassOptionsToPackage{prologue,dvipsnames,table}{xcolor}
\documentclass[manuscript]{acmart}

\usepackage{wrapfig}
\usepackage{minted}
\usepackage{float}
\usepackage{subcaption}
\usepackage{booktabs}
\usepackage{multirow}
\usepackage{makecell}
\usepackage{colortbl}
\usepackage{pgf} 
\usepackage{highlight}
\usepackage{etc}
\usepackage{array}

\newminted{java}{
  frame=lines,      
  framesep=2mm,     
  baselinestretch=1.2, 
  fontsize=\small,  
  linenos,          
  breaklines        
}

\AtBeginDocument{%
  }

\setcopyright{none}
\copyrightyear{2026}
\acmYear{2026}
\acmDOI{XXXXXXX.XXXXXXX}
\acmConference[Submitted to TOSEM]{}{2026}{}

\begin{document}
\raggedbottom

\title[\textit{Efficient Interns or Senior Engineers}: On LLMs’ Grasp of Cohesion and Coupling]{\textit{Efficient Interns or Senior Engineers}: On LLMs’ Grasp of Cohesion and Coupling}

\author{Mootez Saad}
\orcid{0009-0008-8159-3632}
\affiliation{%
  \institution{Dalhouise University}
  \city{Halifax}
  \country{Canada}}
\email{mootez@dal.ca}

\author{Boqi Chen}
\orcid{0000-0002-1451-3603}
\affiliation{%
  \institution{University of Ottawa}
  \city{Ottawa}
  \country{Canada}}
\email{boqi.chen@uottawa.ca}

\author{Jos\'e Antonio Hern\'andez L\'opez}
\orcid{0000-0003-2439-2136}
\affiliation{
  \institution{University of Murcia}
  \city{Murcia}
  \country{Spain}}
\email{joseantonio.hernandez6@um.es}

\author{D\'aniel Varr\'o}
\orcid{0000-0002-8790-252X}
\affiliation{%
  \institution{Link\"oping University}
  \city{Link\"oping}
  \country{Sweden}}
\email{daniel.varro@liu.se}

\author{Tushar Sharma}
\orcid{0000-0002-0538-052X}
\affiliation{%
  \institution{Dalhouise University}
  \city{Halifax}
  \country{Canada}}
\email{tushar@dal.ca}

\renewcommand{\shortauthors}{Saad et al.}

\begin{abstract}
Large language models (\llm{}s) are being increasingly adopted in the software engineering domain, yet the robustness of their grasp on core software design concepts remains unclear. We conduct an empirical study to systematically evaluate their understanding of cohesion (intra-module) and coupling (inter-module). We programmatically generate poorly designed code fragments and test \added{three} DeepSeek-R1 models ($14$B, $32$B, $70$B) \added{along with two proprietary reasoning models, GPT-$5$ Nano and Gemini-$2.5$ Flash Lite,} under varying levels of guidance, from simple \textit{Verification} to \textit{Guided} and \textit{Open-ended Generation}, while varying contextual noise by injecting distractor elements. While models exhibit a solid baseline understanding of both concepts in ideal conditions, their practical knowledge is fragile and highly asymmetrical. Reasoning about coupling proves brittle; performance collapses in noisy, open-ended scenarios, with F1 scores dropping by over $50\%$. In contrast, the models' analysis of cohesion is remarkably robust to internal noise in guided tasks, showing little performance degradation. However, this resilience also fails when all guidance is removed. 
\added{We also conduct a human study involving experienced developers and researchers to contextualize the performance of said \llm{}s. Results show that the human-\llm{} gap is small on \textit{Verification} but widens as guidance is reduced and noise is increased.}
Reasoning-trace analysis confirms these failure modes, revealing \textit{cognitive shortcutting} for coupling versus a more exhaustive (yet still failing) analysis for cohesion for \added{open-weight models}. To summarize, while \llm{}s can provide reliable assistance for recognizing design flaws, their ability to reason autonomously in noisy, realistic contexts is limited, highlighting the critical need for more scalable and robust program understanding capabilities.
\end{abstract}



\maketitle
\section{Introduction}
Large Language Models (\llm{}s) are becoming increasingly embedded in modern software engineering workflows. From automated code generation~\cite{Jiang2024survey} to bug localization~\cite{Hossain2024deep}, models such as GPT-$4~$~\cite{Achiam2023} and DeepSeek-R$1$~\cite{Guo2025} have shown remarkable proficiency in assisting developers in multiple tasks. Existing research has mainly focused on assessing whether \llm{}s can generate functionally correct code, that is, code that compiles and passes predefined test cases~\cite{Chen2021CODEX, Austin2021program, Jimenez2024, Zhuo2024BCB, Jain2024LCB}. However, as projects scale, the long-term success of a codebase depends on its design quality~\cite{Basili1996, Palomba2018}.

Recent works showed that \llm{}-generated code often suffers from poor maintainability, fails to adhere to design best practices, and can introduce structural flaws during refactoring~\cite{Pudari2023, Liu2024empirical, Cordeiro2024, Molison2025}. These studies document the \textit{symptoms} of poor design, but they do not diagnose the root cause. This leaves two critical questions unanswered: First, do \llm{}s possess a core, foundational knowledge of design principles? Second, \textbf{how robust is that knowledge}? Does it only surface when a problem is presented as a simple, highly-guided task, or can the model apply these principles autonomously in complex, open-ended scenarios?

This leads us to the central analogy of our work: \textit{are \llm{}s akin to efficient interns, capable of flawlessly executing well-defined tasks specified using prompts, or do they exhibit the analytical autonomy of senior software engineers, who can navigate ambiguity and natively adopt best-practices derived from design principles?}

The goal of the study is to investigate the knowledge that language models of code have about software design and the conditions under which they can manipulate and apply it.
Specifically, we focus on two foundational software design principles: \textit{cohesion} and \textit{coupling}. These principles are important to create maintainable and adaptable software. Cohesion describes how closely the elements within a module relate to one another~\cite{Yourdon1979}, while coupling characterizes the degree of interdependence between different modules~\cite{Bourque2014}.
Despite their importance, little empirical evidence exists about whether \llm{}s can recognize or reason about them. 

Our study design draws inspiration from Bloom's Taxonomy~\cite{Bloom1964} to achieve the above-stated goal.
It is a framework for categorizing cognitive skills to investigate the cognitive depth of human knowledge. In software engineering literature, it has been used by Buckley and Exton~\cite{Buckley2003} to evaluate the developers' comprehension of a software system.
We devise an experimental framework that operationalizes its levels (such as \textit{understanding}, \textit{evaluating}, and \textit{creating})
through varying the instruction type and provided contextual information in our prompts to \llms{}.
Specifically, we design prompts that exhibit decreasing levels of guidance: (1) \textbf{Verification}: the model is presented with an assertion and must simply confirm or deny it, (2) \textbf{Guided generation}: the model is given a partial context and must apply its knowledge to complete the task; and (3) \textbf{Open-ended generation}: the model is given only the raw code and must autonomously discover design elements.

Our evaluation framework evaluates three variants of the DeepSeek-R$1$ model ($14$B, $32$B and $70$B) \added{along with two proprietary reasoning models, GPT-$5$ Nano and Gemini-$2.5$ Flash Lite,} and controls for two key variables: the level of contextual help provided, and by extension, the \textit{cognitive load} and the \textit{distortion levels}, which introduce controlled noise into the input space to mimic a realistic development environment. To introduce controlled noise, for cohesion, we inject methods from unrelated classes into highly cohesive ones; for coupling, we introduce distractor classes around truly coupled pairs that were generated using a set of code transformations. This structure allows us to quantify whether model performance is consistent and aligned with software engineering principles under real constraints, leading us to our research questions:
\begin{itemize}
\item[\textbf{RQ1:}] \textit{What is the \llm{}'s baseline performance in recognizing the concepts of cohesion and coupling}? \newline
This question establishes the model's baseline \textbf{capability}. Our goal is to measure its core competency in distinguishing between cohesive and non-cohesive classes, and between tightly and loosely coupled classes in a controlled, low-noise environment. This serves as a reference point for all subsequent questions. Our results show that \llm{}s recognize cohesion and coupling in simple tasks, but performance drops with complexity. The $70$B model scores $0.825$ (cohesion) and $0.899$ (coupling) in ideal conditions, yet falls to $0.649$ and $0.736$ in complex guided generation tasks. \added{The proprietary models reach $\sim0.96$ F1 on coupling \textit{Verification}, exceeding the $70$B open-weight model, but this lead does not carry over to the more autonomous settings (\textit{Open-ended Generation}).}

\item[\textbf{RQ2:}] \textit{To what extent does the level of contextual help in a prompt influence the model's performance?} \newline
Building on the baseline established in RQ1, we next probe the \textbf{autonomy} of the model's knowledge. We measure performance across different levels of help: \textit{Verification} (maximum), \textit{Guided Generation} (partial), and \textit{Open-ended Generation} (minimal). A performance drop on open-ended tasks would suggest the model excels as an executor of instructions but struggles with independent discovery. Model performance declines as cognitive load increases \added{across both open-weight and proprietary families}, showing strong reliance on guidance. For coupling, the $70$B model drops over $31\%$ (F1: $0.89 \rightarrow 0.61$), while cohesion reaches an ARI score of $0.43$ in open-ended tasks.\added{The drop is steepest for GPT-$5$ Nano, whose coupling F1 falls from $0.918$ in \textit{Verification} to $0.230$ in \textit{Open-ended Generation}.}

\item[\textbf{RQ3:}] \textit{How does increasing distortion level impact model performance}? \newline
Whereas the previous questions assess performance in ideal settings, this question stress-tests the \textbf{robustness} of the model's knowledge against noise. We systematically inject irrelevant distractor code to simulate a more realistic environment and identify the model's breaking points. This allows us to determine if the model's reasoning is fragile when noise increases. Models' performance on coupling tasks declines with distortion, with the $70$B model dropping from $0.59$ to $0.27$ F1 in open-ended tasks and similar trends across smaller models. On cohesion tasks, models remain stable in verification and guided settings (\eg{} $0.825$ F1 for the $70$B model), but degrade in open-ended scenarios (ARI: $0.44 \rightarrow 0.25$).

\added{This pattern holds across both open-weight and proprietary model families. The proprietary models retain higher coupling \textit{Verification} accuracy under high distortion and sustain higher ARI scores on open-ended cohesion than every open-weight variant, though coupling remains the more fragile concept for every model.}

\item[\added{\textbf{RQ4:}}] \added{\textit{How does \llm{} performance on cohesion and coupling tasks compare to that of human developers?}} \newline
\added{While RQ1–RQ3 characterize the models' capabilities in isolation, this question grounds those results against a human baseline. Comparing models to experienced developers under increasing cognitive load and distortion reveals how their performance gap changes. The human-\llm{} gap is small on \textit{Verification} but widens as guidance is removed and noise increases. At the hardest setting, human developers hold their largest leads ($0.483$ vs.\ $0.377$ for the best \llm{} on coupling, and $0.704$ vs.\ $0.419$ on cohesion), confirming the intern-level profile of the evaluated models.}

\item[\textbf{RQ\added{5}:}] \textit{What are the main trends in reasoning traces generated by the models under the studied experimental conditions}?  \newline
Through this investigation, we analyze the trends of the chain-of-thought traces (CoT) generated by the models before their final answer. Our goal is to investigate how they differ across different experimental conditions and the possible reasons behind such differences. In coupling tasks, surprisingly, \added{the open-weight} \llm{}s produce shorter reasoning traces in open-ended settings than in guided ones. For cohesion, trace length grows with cognitive load, particularly in open-ended settings. Across both concepts, \added{open-weight} larger models ($32$B, $70$B) generate shorter traces than the smaller $14$B model.
\end{itemize}

The contributions of this work are as follows:
\begin{itemize}
    \item \textbf{A novel benchmark and methodology for the controlled evaluation of \llm{} reasoning on two core software design concepts}: our approach enables the systematic generation of code with quantifiable cohesion and coupling flaws, facilitating reproducible experiments on model robustness.
    \item \textbf{The first in-depth empirical study revealing an asymmetry in how models handle cohesion versus coupling}: we provide quantitative evidence that while models possess foundational knowledge, their practical reasoning about coupling is highly susceptible to noise, whereas their cohesion analysis is more robust but brittle under full autonomy. \added{This asymmetry holds across both open-weight and proprietary families.}
    \item \added{\textbf{A human baseline that grounds model performance against a real-world reference}: experienced software developers and researchers complete the same cohesion and coupling tasks under the same experimental conditions, which lets us measure the human-\llm{} gap across cognitive loads and distortion levels.}
    \item \textbf{Mechanistic insights into the failure modes of \llm{}s for design analysis, derived from Chain-of-Thought traces \added{produced by open-weight models}}: we identify and contrast distinct reasoning patterns: \textit{cognitive shortcutting} when faced with combinatorial complexity in coupling tasks versus a \textit{brute-force} analysis for cohesion.
    \item \textbf{Actionable insights for researchers and practitioners on the readiness of \llm{}s for software design tasks \added{related to coupling and cohesion}}: our findings characterize current models as effective guided assistants for identifying suboptimal design but caution against their use as fully autonomous agents in noisy, complex codebases.

\end{itemize}
Our replication package, including source code and data, is available online~\cite{replication}.
\section{Related Work}

\subsection{Nonfunctional Requirements of Software}
Functional correctness is insufficient for capturing the spectrum of what it means to produce \textit{high-quality} software. Real-world code should not only be correct, but also be maintainable and well-structured. Maintainability is a key quality characteristic of software products, which can be decomposed into five sub-characteristics: \textit{modularity}, \textit{reusability}, \textit{analyzability}, \textit{modifiability}, and \textit{testability}~\cite{iso25010}. These attributes are directly influenced by how well a codebase adheres to established design principles dictated by concepts such as cohesion and coupling. 

However, multiple works have shown that the output of \llm{}s often suffers from inconsistencies that hinder maintainability~\cite{sonarsource2024, turintech2024, devops2024}. Molison~\etal{}~\cite{Molison2025} found that \llm{}-generated code can introduce \textit{critical structural issues} not present in human-written code, despite offering efficiency in resolving bugs. Another empirical evaluation by Nunes~\etal{}~\cite{Nunes2025} showed that even when language models correctly identify maintainability issues, the majority of their fixes lead to compilation errors or regressions.

Further, recent taxonomies of \llm{}-generated inefficiencies~\cite{Abbassi2025} highlight quality concerns such as redundancy, suboptimal performance, and the need for extensive post-generation human oversight. As the adoption of code-generating \llm{}s continues to rise~\cite{Kelly2024}, evaluating their capacity to uphold structural integrity is important.

\subsection{Software Design Capabilities of Large Language Models}

Pudari and Ernst~\cite{Pudari2023} provide further empirical evidence of this gap by exploring the limitations of AI-supported code completion tools. They evaluated GitHub Copilot's ability to adhere to established Python language idioms and to avoid common code smells in JavaScript. Their findings showed that Copilot failed to generate the preferred code in its primary suggestion; it did not suggest the idiomatic Python approach in $92$\% of test cases and failed to produce the non-smelly JavaScript best practice in $88$\% of cases. To frame these limitations, they proposed a hierarchy of software abstractions, with \textit{syntax} and \textit{correctness} at the bottom and more complex concerns like \textit{paradigms and idioms}, \textit{code smells}, and \textit{design level} at the top. They conclude that while current tools can achieve functional correctness, they are far from being able to handle the more abstract principles of software engineering that are essential for quality and maintainability.

Similarly, Liu et \etal{}~\cite{Liu2024empirical} conducted an empirical study on the capabilities of ChatGPT and Gemini in automated software refactoring. They investigated two core tasks: identifying refactoring opportunities and recommending refactoring solutions, using a dataset of $180$ real-world refactorings. When using generic prompts, both models performed poorly at identifying opportunities, with success rates of only $15.6$\% for ChatGPT and $3.9$\% for Gemini. However, the study demonstrated that performance could be improved through prompt engineering by specifying refactoring subcategories and narrowing the code's search space. Doing so increased ChatGPT's success rate to $86.7$\%. In evaluating the quality of recommended solutions, they found that over $60$\% of ChatGPT's suggestions were comparable to or better than those from human experts. Crucially, however, they also identified a significant safety risk, as $7.4$\% of ChatGPT's solutions and $6.6$\% of Gemini's introduced bugs, either by changing the code's functionality or causing syntax errors. 

Recently, research by Cordeiro \etal{}~\cite{Cordeiro2024} into automated software refactoring reveals the challenges \llm{}s face with behavior-preserving transformations. While their study showed that these models could successfully execute simple, localized refactorings like \textit{rename variable}, they struggled significantly with more complex, structural operations. A substantial portion of the generated refactorings were semantically flawed, either introducing compilation errors and test failures or failing to achieve the intended design improvement. The authors noted that the models often performed the refactoring mechanically, without a deeper comprehension of the underlying design principles, sometimes even degrading the code's structure.

Recent works have also examined \llm{}s in the context of code smells and design patterns, two important software design concepts. Silva~\etal{}~\cite{Silva2024} evaluated \texttt{gpt-3.5-turbo} on its ability to detect four code smells: Blob, Data Class, Feature Envy, and Long Method. They used two prompts, where the first is a generic one that asks the model to identify any smell in a Java snippet, whereas the second prompt explicitly named the four targeted smells. Their results indicated how the more specific prompt yielded a higher F1 score across all code smells and that detection improved with smell severity. Along similar lines, Liang~\etal{}~\cite{Liang2025} introduced \textit{SmellDetector}, a multi-label classifier built by fine-tuning CodeLlama-$7$B with QLoRA adapters on a curated dataset of $\sim212K$ instances across $20$ class- and method-level smells. They further showed that smell information, when integrated as auxiliary input, can improve performance on a refactoring-opportunity detection task.

In addition, some studies recently evaluated \llm{}s' capabilities to detect and introduce design pattern instances in code.
Specifically, Pandey~\etal{}~\cite{Pandey2025} evaluated five language models, Code2Vec, CodeBERT, CodeGPT, CodeT5 and RoBERTA, on their ability to recognize five Gang-of-Four design patterns in Java code. Their methodology consisted of extracting frozen embeddings from each model and training a $k$-NN classifier to label instances of a given pattern. Kim~\cite{Kim2025} extended the focus from recognition to application. They evaluated four commercial models (ChatGPT, Claude, Copilot, and Gemini) alongside the open-weight Llama-$3.1$-$405$B model on whether they can correctly apply five design patterns to Java systems. The generated refactoring was evaluated against predicate-logic design pattern specifications through three metrics: Property Satisfaction Rate, Critical Property Coverage, and Pattern Implementation Quality Score. Similar to the work of Silva~\etal{}~\cite{Silva2024}, they have also used two different kinds of prompts that differ in terms of explicitness.

\subsection{Limitations of Existing Work}
Despite the breadth of research, these works highlight a gap in the capabilities of current \llm{}s. While models can generate functionally correct code, they often falter when it comes to the principles of software quality that ensure long-term maintainability. Existing research has documented the \textit{symptoms} of this deficiency from multiple angles: demonstrating that language models struggle to adhere to language idioms, avoid code smells~\cite{Pudari2023}, or perform safe and meaningful refactorings without significant human guidance and post-processing~\cite{Liu2024empirical}, and evaluating how well they detect code smells~\cite{Silva2024, Liang2025} or recognize and apply
established design patterns~\cite{Pandey2025, Kim2025}. However, what remains unaddressed is a systematic evaluation of the models' understanding of the \textit{foundational design principles} that underpin these practices. The literature has yet to measure an \llm{}'s core grasp of fundamental concepts like software cohesion (an intra-module concern) and coupling (an inter-module concern). The importance of these concepts was underscored in a study by Wan~\etal{}~\cite{Wan2023SW}. Through interviews with 32 participants from 21 organizations spanning three continents, they found these concepts to be recurring topics of importance among software architects.
This study aims to fill this gap by directly evaluating the models' knowledge of these principles, thereby moving from observing the consequences of poor design to \underline{\textit{diagnosing the models' core comprehension of software structure itself}}. Furthermore, it aims to understand the \textit{conditions} under which this understanding holds or degrades by varying the amount of guidance provided to the models, and the contextual noise. Finally, it seeks to contextualize the performance of such models against that of human developers.
\section{Methodology}
\begin{figure}[h!]
    \centering
    \includegraphics[width=\linewidth]{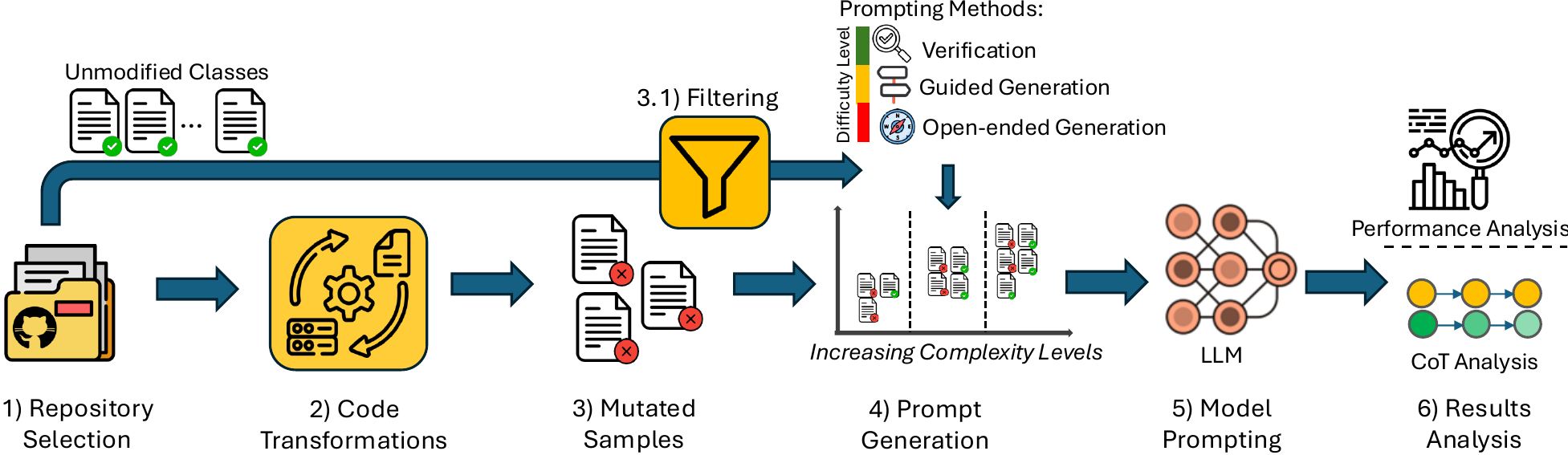}
    \caption{Overview of the methodology. 
    }
    \label{fig:overview_figure}
\end{figure}

In this section, we describe the methodology used to assess the understanding of \llm{}s of coupling and cohesion.
As illustrated in Figure~\ref{fig:overview_figure}, we first describe our data collection process and the set of code transformations to generate code fragments where coupling and cohesion are manifested. Then, we describe the prompt construction process. 
Finally, we present the constructed prompts to the selected models and analyze the obtained results.

\subsection{Raw Data Collection}
In this step, we collect a list of open-source Java projects that will be used to generate samples related to cohesion and coupling. We focus on Java given its object-oriented nature, which makes it ideal for studying coupling and cohesion in addition to its wide adoption~\cite{StackOverflow2025Survey, TIOBE2025Sep}. We use the SEART tool~\cite{Dabic2021} to select repositories that meet the following criteria: a star count of $1K$ or more as a proxy of usefulness, a minimum of $1K$ commits as a filter of maturity, non-forked, have a maximum of $150K$ lines of code to keep the processing times reasonable, and actively maintained by specifying the latest commit date to be of the month during which the study is conducted, as done in similar works~\cite{Avelino2019, Zerouali2025}. This resulted in a set of 219 repositories. We discard the repositories that have build/compilation issues yielding final list of $210$ repositories.

\subsection{Code Transformations}
\label{sbs:code_transformations}

To systematically evaluate \llms{} in handling degraded code quality, we define code transformations that generate samples with \textbf{controlled} and \textbf{verifiable} manifestations of \textit{high coupling} and \textit{low cohesion}, at scale. This control is essential because the natural, heterogeneous occurrence of these properties in real projects complicates targeted and systematic assessment. 
\begin{figure}[H]
        \includegraphics[width=.6\linewidth]{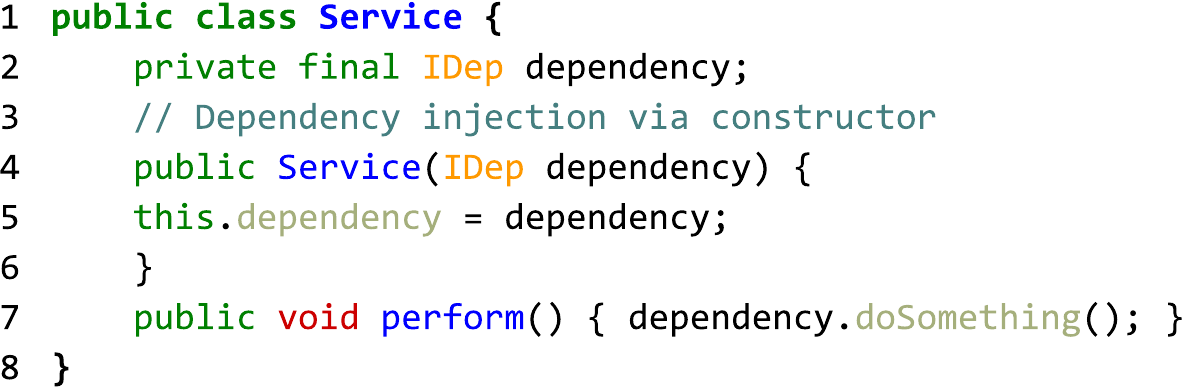}
        \caption{Original code snippet before transformation.}
        \label{fig:original_code}
\end{figure}

Crucially, we also aim to reduce the threat of \textit{data contamination} by generating realistic data absent from the models' training corpora. 
These transformations are implemented using the Spoon~\cite{Pawlak2015} library. For clarity, we provide the output of each transformation when applied to the code snippet shown in Figure~\ref{fig:original_code}.

\subsubsection{Direct Instantiation of Dependencies - \did{} (Coupling)}
This transformation introduces tighter coupling between Java classes that implement dependency injection~\cite{Fowler2004injection} and their corresponding dependencies.
Using Yang~\etal{}'s criteria~\cite{Yang2008}, we identify dependency injection sites where \textit{all} parameters are exclusively interface or abstract class types to minimize false positives.

\begin{figure}[H]
        \includegraphics[width=.6\linewidth]{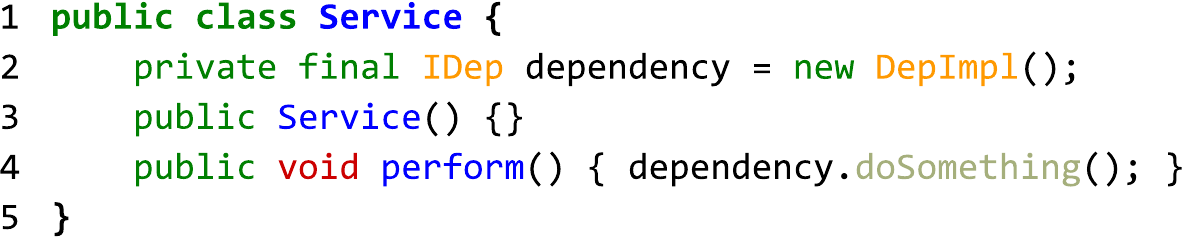}
        \caption{Code snippet after the \did{} transformation.}
        \label{fig:did_transformation}
\end{figure}

For each identified location, we then:
\begin{itemize}
    \item remove all parameters from the method signature,
    \item select concrete implementations randomly for each interface or abstract class, and
    \item replace dependency assignments with direct field instantiations using these implementations.
\end{itemize}
The generated code introduces tighter coupling between a class and its dependencies through hard-coded instantiation.

\subsubsection{Unused Injected Dependency - \uidep{} (Coupling)}

The following transformation is similar to \did{}. The difference is that we keep the method signature intact, but we directly instantiate the dependency within the method's body. It is less explicit and provides fewer cues to recognize tight coupling. 

\begin{figure}[H]
        \includegraphics[width=.6\linewidth]{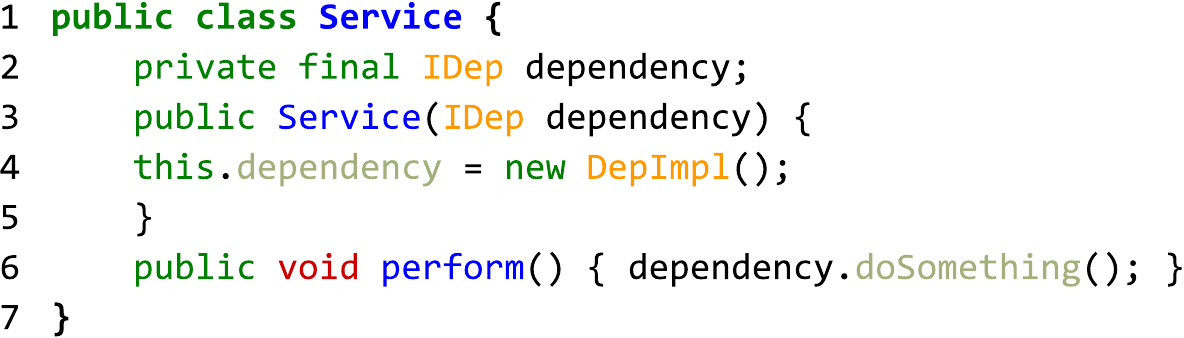}
        \caption{Code snippet after the \uidep{} transformation.}
        \label{fig:uid_transformation}
\end{figure}

\subsubsection{Indirect Instantiation of Dependencies - {\textsc{idd}} (Coupling)}

{\textsc{idd}} introduces an intermediary factory-like class to obscure the direct instantiation of a dependency. The consuming class is modified to call a static factory method that was injected in the dependency class, which, in turn, is responsible for instantiating and returning the dependency object. 

\begin{figure}[H]
        \includegraphics[width=.6\linewidth]{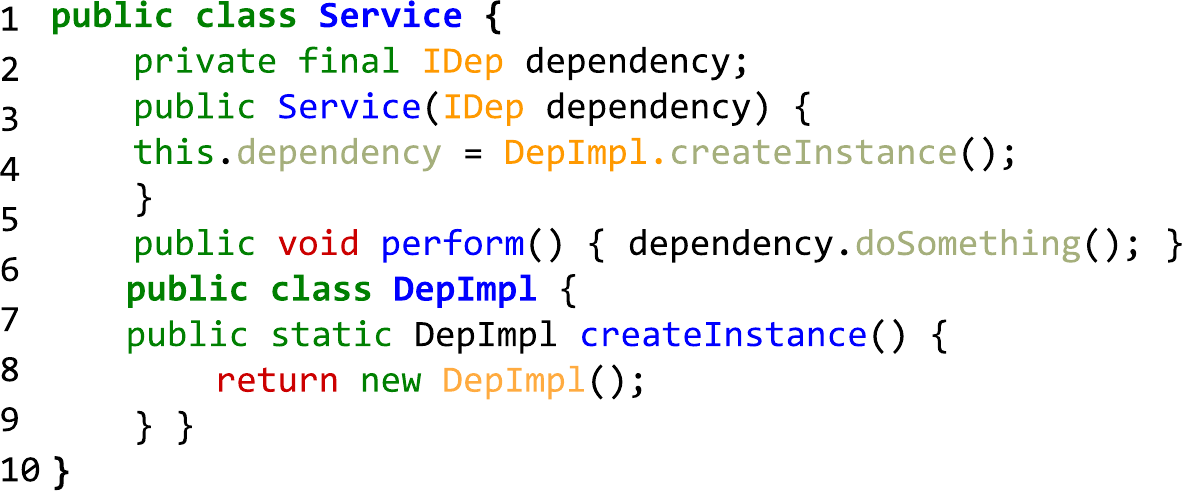}
        \caption{Code snippet after the {\textsc{idd}} transformation.}
        \label{fig:iid_transformation}
\end{figure}

Although this abstracts the direct use of the \texttt{\textbf{new}} keyword within the consuming class, it still establishes tight coupling by making the consumer dependent on the concrete factory class rather than relying on an injected interface.

As a final step, across all transformations applied for coupling experiments, we \textit{skeletonize} the classes, \ie{} keep only method signatures (excluding constructors) and remove comments,  before presenting them to the model. We do so to ensure that the manifestation of tight coupling is only the result of our code transformation. If method bodies were left intact, we could not guarantee the absence of tight coupling between a pair of classes. 
For instance, a non-constructor method in class X can directly instantiate class Y, thereby introducing tight coupling between X and Y. Hence, by following such a strategy, we objectively know which pairs of classes are tightly coupled and which are not. In addition, the scope of this study is structural coupling. There may be other manifestations of other types of coupling within such methods that would lead to noisy labeling. Furthermore, we want to keep the input length within the context window of the models under test.

\subsubsection{Introducing In-cohesive Class (Cohesion)} 
This transformation is structured into two main phases: identifying suitable cohesive classes and a controlled injection transformation. 

\begin{figure}[h!]
    \centering
    \includegraphics[width=\linewidth]{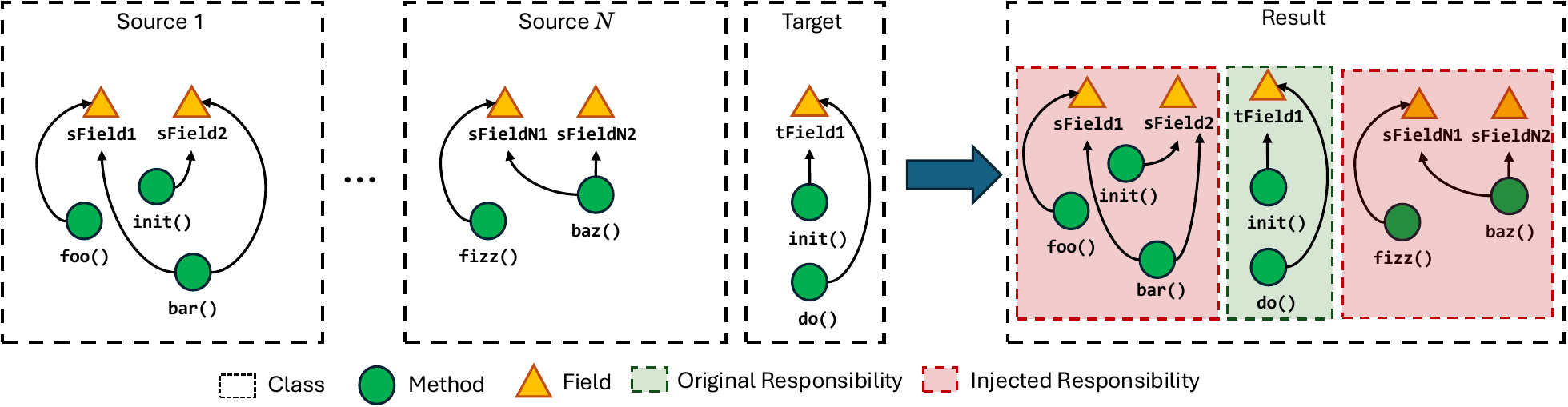}
    \caption{An overview of the transformation logic for producing loosely cohesive classes.}
    \label{fig:cohesion_transformation}
\end{figure}

To identify the pool of these cohesive classes, we consider those with a YALCOM~\cite{Sharma2020} value of $0$ and a total method count greater than two. YALCOM is a lack-of-cohesion metric, an improvement over traditionally used LCOM4~\cite{Hitz1996_LCOM4} metric by incorporating, for example, method calls, in calculating the metric where higher values indicate a greater degree of disconnection among a class's methods and fields. We refer the reader to Sharma~\etal{}~\cite{Sharma2020} for its formal definition. We use the implementation provided by DesigniteJava~\cite{Sharma2016} which has been used in previous works~\cite{Cordeiro2024, Arcelli2023, Horikawa2025, horikawa2025agentic}.
We adopt this heuristic as it has been used in code smell detection to avoid generating false-positive instances of the \textit{multifaceted-abstraction smell}~\cite{Sharma2020b}. Using this pool, we generate incohesive artifacts through a controlled injection process. We select a \textit{target} class and a set of \textit{source} classes. We ensure that the methods from a certain class do not call or share fields with other methods from another class. Then, $N$ methods are randomly selected for each source class. To maintain the structural integrity of the injected logic, any class fields from a source class that are accessed by these methods are identified for transfer. Finally, a new class is created by cloning the target class and then injecting the $N$ selected methods and their dependent fields from the source classes. This process is illustrated in Figure~\ref{fig:cohesion_transformation}.

To eliminate potential confounding variables, two final controls are enforced. First, the newly generated class is given a generic, randomized name (\eg{} \texttt{GeneratedClass1a2b3c4d}). This prevents the class name from providing any hints to the \llm{} about its composite nature, forcing the model to reason about the code's structure rather than its name. Second, we only pair source and target classes that do not share any common, non-\texttt{Object}\footnote{\texttt{Object} as in the class Object that is the root of the class hierarchy in Java.} ancestor in their inheritance hierarchies. This mitigates the risk of shared inherited behavior confounding the model's assessment of cohesion.
Additionally, we exclude test classes during data generation for the coupling and cohesion experiments.

\subsubsection{Validity of Generated Code Snippets} Dependency Injection is a design pattern that implements the \textit{Inversion of Control} (IoC) principle to mitigate tight coupling between software modules~\cite{Fowler2004injection}. 
By delegating the responsibility of object creation, {\sc{DI}} ensures that a class receives its dependencies from an outside source rather than instantiating them directly. This separation of concerns compels components to depend on abstractions, such as interfaces, rather than concrete implementations. Consequently, the system achieves a loosely coupled architecture where modules are highly modular and testable, as dependencies can be seamlessly interchanged or mocked without requiring modifications to the client code. Hence, inverting this design pattern by making classes depend on \textit{concrete} implementations introduces tighter coupling between classes. As a result, these generated code elements are valid by \textbf{construction}.

\begin{figure}[H]
    \centering
    \includegraphics[width=0.4\linewidth]{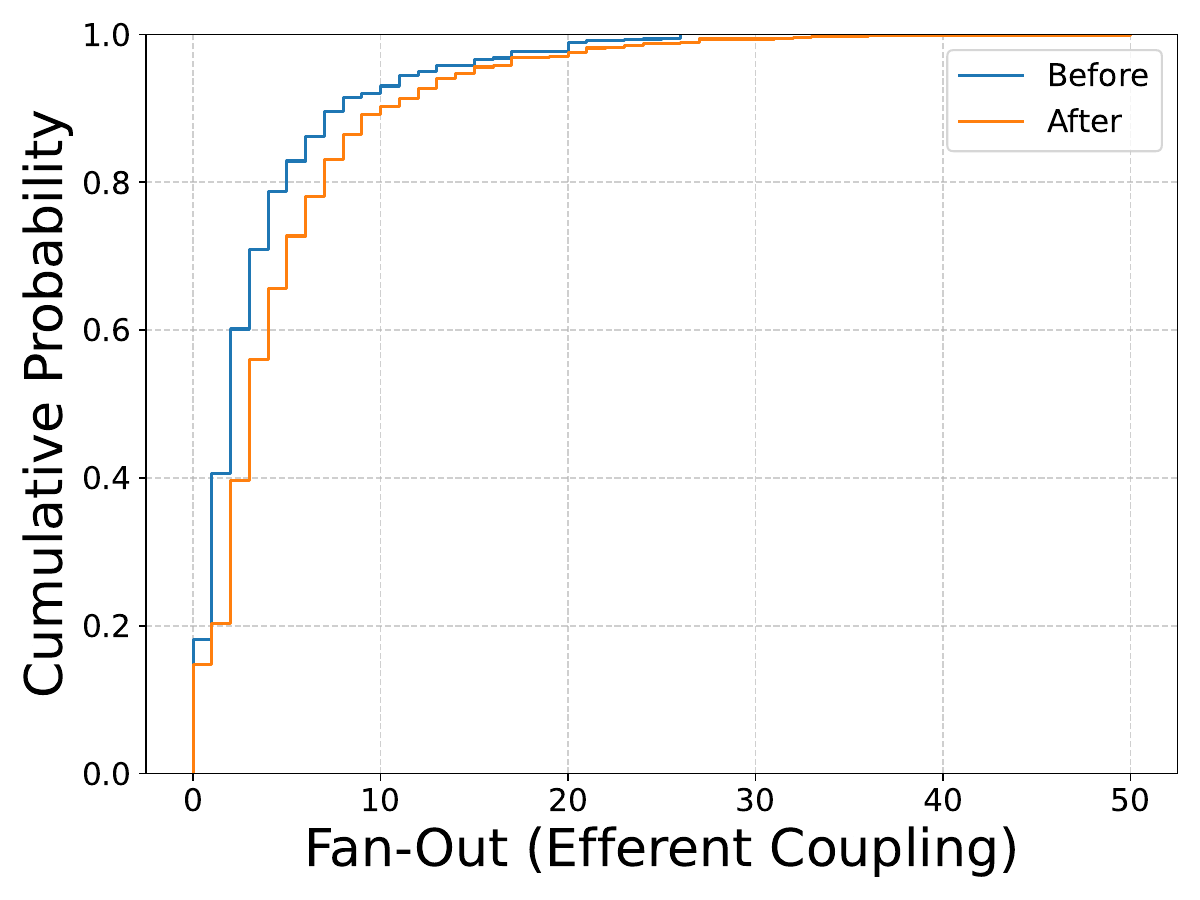}
    \caption{eCDF of Fan-Out scores, illustrating the changes in external dependencies and class coupling resulting from the code transformations. These changes are manifested through the rightward shift of the curve after the code transformation is applied.}
    \label{fig:ecdf_coupling}
\end{figure}

We provide empirical evidence to further validate the generated code elements. We compare the \textit{fan-out} metric of each modified class before and after the transformation is applied. Fan-out counts the number of distinct classes on which a given class depends, and is a well-established structural proxy for efferent coupling~\cite{martin1994oodesign}. Given that the code transformations replace dependencies with concrete implementations, an increase in tight coupling should manifest as a rightward shift in the distribution of transformed classes. Figure~\ref{fig:ecdf_coupling} reports the empirical cumulative distribution functions (eCDFs) of fan-out values aggregated across the three transformations.

The post-transformation eCDF lies consistently to the right of that of the pre-transformation, indicating that transformed classes depend on a larger number of concrete types. A paired Wilcoxon signed-rank test confirms that this shift is statistically significant ($p < 0.001$), and Cliff's Delta ($\delta = -0.22$) indicates a small but non-negligible effect size~\cite{Romano2006exploring}. The modest effect size is consistent with the localized nature of the transformations: each transformation alters only the dependency injection sites of a class rather than its overall structure, so the change in fan-out is bounded.

For cohesion, we enforce low cohesion by verifying that the graph of a class, where its methods and attributes are the vertices and edges represent field access and invocations, contains exactly \textit{N+1} disconnected subgraphs transferred from unrelated classes, objectively violating the Single Responsibility Principle. Therefore, the resulting class includes clusters of methods with non-overlapping responsibilities.

Since the code transformation is parameterized by the number of source classes $N$ from which unrelated methods are injected, we plot the empirical cumulative distribution functions of the YALCOM values of the Java classes post-transformation when varying such parameter between $1$ and $9$.
\begin{figure}[H]
    \centering
    \includegraphics[width=0.5\linewidth]{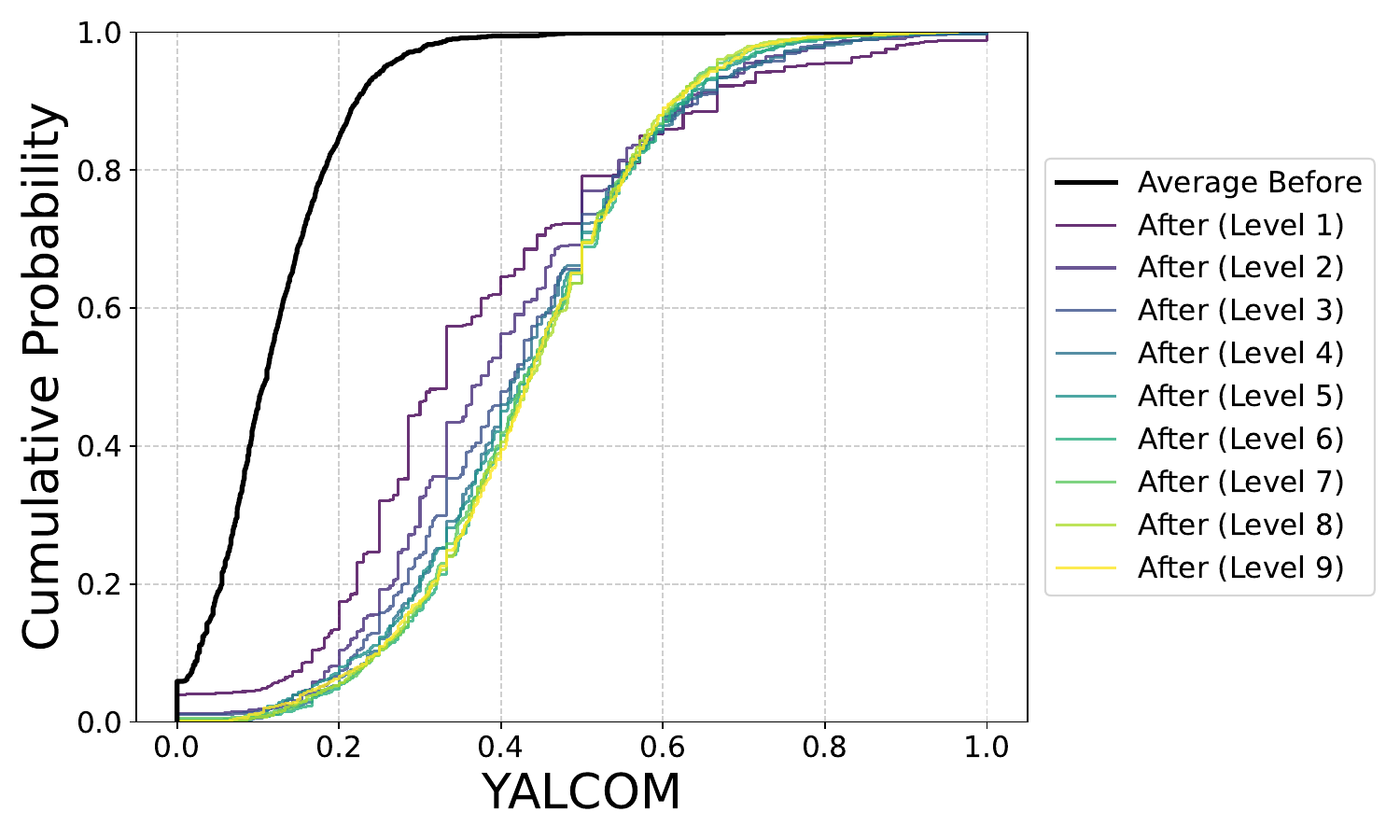}
    \caption{eCDF comparing the baseline (Before) YALCOM scores against the generated (After) classes. A shift in the distribution indicates the degree to which class cohesion was altered by the transformation.} 
    \label{fig:ecdf_cohesion}
\end{figure}
YALCOM is a lack-of-cohesion metric where higher values indicate a greater degree of disconnection among a class's methods and fields, so a genuine reduction in cohesion should manifest as a rightward shift in the YALCOM distribution of transformed classes. Indeed, this is what is observed in Figure~\ref{fig:ecdf_cohesion}. For every $N$, the post-transformation eCDF lies consistently to the right of the pre-transformation eCDF.
\begin{table}[H]
    \caption{\added{Statistical analysis of YALCOM variations when varying the number of source classes. The table reports Wilcoxon signed-rank test results comparing baseline class cohesion against generated equivalents. Significance ($\alpha = 0.05$) is evaluated using Benjamini-Hochberg adjusted p-values to control the false discovery rate across multiple comparisons. The first column represent the number of source classes. Effect sizes are quantified using Cliff's Delta ($\delta$).}}
    \centering
    \label{tab:cohesion_stat_analysis}
\resizebox{.65\textwidth}{!}{
    \begin{tabular}{lrrrrrl}
    \toprule
    \makecell{\#Classes} & Adjusted $p$-value & Cliff's Delta & Magnitude \\
    \midrule
    1 & $< 0.001$  & -0.609 & large \\
    2 & $< 0.001$  & -0.697 & large \\
    3 & $< 0.001$  & -0.726 & large \\
    4 & $< 0.001$  & -0.760 & large \\
    5 & $< 0.001$  & -0.771 & large \\
    6 & $< 0.001$  & -0.794 & large \\
    7 & $< 0.001$  & -0.797 & large \\
    8 & $< 0.001$  & -0.810 & large \\
    9 & $< 0.001$  & -0.802 & large \\
    \bottomrule
    \end{tabular}%
}
\end{table}

In Table~\ref{tab:cohesion_stat_analysis}, we report the results of the Wilcoxon signed-rank test, with Benjamini-Hochberg correction to control the false discovery rate across the nine comparisons, along with the effect sizes. The results confirm that every shift is statistically significant ($p < 0.001$). Furthermore, Cliff's Delta indicates a large effect size~\cite{Romano2006exploring} in all configurations, with the magnitude $|\delta|$ increasing monotonically from $\sim0.61$ at $N=1$ to $0.81$ at $N=8$ before plateauing at $N=9$. This illustrates how the degradation of the cohesion scales with the number of injected responsibilities as intended. As for the plateau at higher values of $N$, it can be explained by a saturation effect, in which additional injected clusters contribute to diminishing incremental disconnection once the class already contains several unrelated responsibility groups.

Finally, although runtime behavior and the execution of code snippets are unnecessary to infer the answers to our benchmark, we take the necessary measures to ensure that the project remains compilable and test suites still pass for the sake of completeness.

\subsection{Prompt Design}
\label{subsec:prompt_design}
To systematically evaluate a model's reasoning capabilities, we designed three prompting strategies, extending similar work~\cite{Kim2025, Silva2024}. These strategies create a \textit{gradient} of cognitive load, ranging from simple recognition to unguided discovery. This tiered approach, inspired by the hierarchical structure of Bloom's Taxonomy~\cite{Bloom1964}, allows us to measure not just if the model understands a concept, but how robustly it can apply that knowledge with varying levels of assistance.

\begin{figure}[H]
    \centering
    \includegraphics[width=\linewidth]{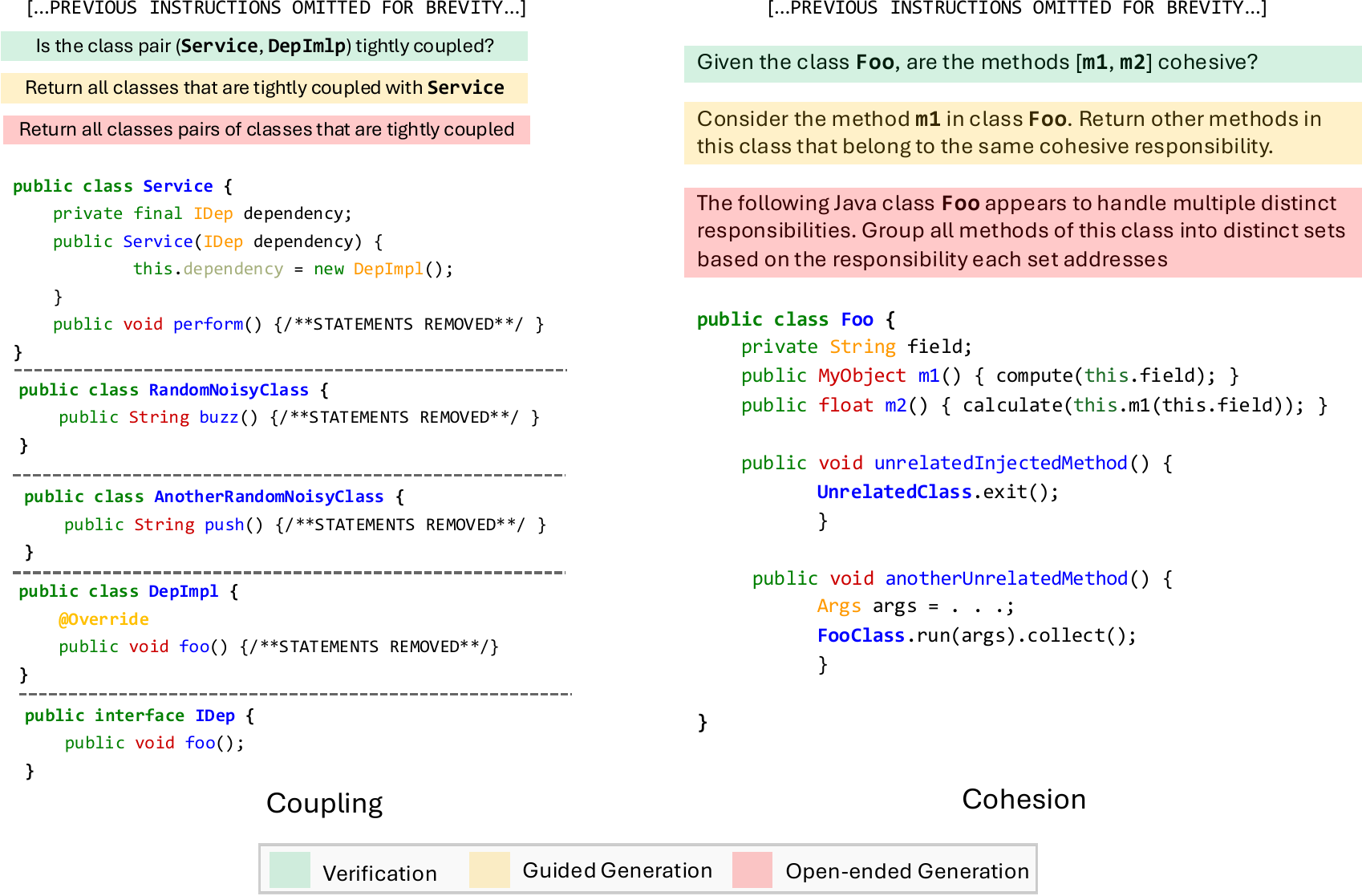}
    \caption{Prompt templates used to carry out the experiments.}
    \label{fig:templates}
\end{figure}

\subsubsection{Verification (Low Cognitive Load)} This represents the lowest level of cognitive demand, testing the model's ability to recognize a design aspect. The model is presented with the source code and a complete, explicit assertion about a design relationship. The model is then required to provide a binary \textit{yes} or \textit{no} response. In coupling, we provide the model with a list of class pairs and ask it whether each pair is tightly coupled or not. In cohesion, the \llm{} is presented with a pair of methods, and is asked whether they are cohesive or not, using this to test whether the model can identify if both methods contribute to the same underlying responsibility. This isolates the model's ability to confirm a fact when it is clearly stated, which is analogous to the \textit{Remembering} and \textit{Understanding} levels of Bloom's Taxonomy. 

\subsubsection{Guided Generation (Medium Cognitive Load)} Moving up the complexity ladder, this task requires the model to apply its knowledge in a constrained context. Instead of a complete assertion, the model is provided with \textit{seed} entities, methods, taken from each cluster, for the cohesion task, or classes that underwent a code transformation, for the coupling task. It is then prompted to identify all other entities in the provided code that are conceptually related to the seeds. In coupling, we provide the model with a class and ask it to return all classes with which it is tightly coupled. Similarly, in cohesion, we provide it with a method and instruct it to return all other methods in this class that belong to the same cohesive responsibility.
This shifts the model from mere recognition to a retrieval-like and analysis task, corresponding to the \textit{Applying} and \textit{Analyzing} levels of Bloom's taxonomy. It measures the model's ability to act on a specific instruction to find related components.

\subsubsection{Open-ended Generation (High Cognitive Load)}
This is the most demanding formulation, which assesses the highest levels of cognitive reasoning. In this scenario, the model is presented only with the source code and is asked to independently discover and partition all relevant conceptual groups (for cohesion) or identify all tightly coupled pairs (for coupling). This task requires the model to first discover the underlying design structure and then evaluate it without any external hints. Success in this task mirrors the \textit{Evaluating} and \textit{Creating} levels of the taxonomy, simulating a realistic refactoring scenario where a developer must identify design flaws from scratch. This formulation is critical for determining if the model possesses a deep understanding of software design principles.

Figure~\ref{fig:templates} illustrates how the models are prompted regarding each design concept under different task formulations. We instruct the model using one guidance level (or cognitive load) at a time \ie{} one prompt contains one guidance level, not the three at the same time.
Furthermore, every code element provided in the context, classes, or methods, is explicitly queried in the instructions for both the \textit{Verification} and \textit{Guided Generation} tasks. However, for the \textit{Verification} level, while every element is guaranteed to be evaluated in at least one pair, we do not prompt the model with the exhaustive list of \textit{all} possible combinatorial pairs.

\subsection{Controlled Distortion Injection}
In addition to varying the level of guidance, we systematically manipulate the \textit{signal-to-noise} ratio of each case to assess the robustness of the model's understanding. Real-world code is seldom clean; classes often contain tangentially related logic, and relevant modules are frequently situated within a larger, noisier project context. Our controlled introduction of noise is designed to simulate this complexity and quantify its impact on the model's performance. 

For cohesion, noise is introduced by directly distorting the cohesion of a class. This is controlled by the number of classes that we choose as the \textit{source}, from which we sample the methods that are injected into the \textit{target} class. This is a direct measure of how much a class's original, single responsibility has been polluted by unrelated concerns. Doing so at controlled distortion levels (between $1$ and $9$) allows us to measure the model's breaking point, the threshold at which it can no longer effectively partition the set of methods into their responsibility sets.

For the coupling tasks, the noise is contextual rather than invasive. The core relationship between the tightly coupled classes remains intact, but it is obscured by introducing independent \textit{distractor} classes into the prompt's context. The \textit{coupling distortion ratio} ($\text{DR}_{coupling}$) is defined as the proportion of these distractor classes relative to the total number of classes presented to the model:

\[
DR_{\text{coupling}} = \frac{\text{Number of Distractor Classes}}{\text{Total Number of Classes}
}
\]

\begin{figure}[H]
    \includegraphics[width=.75\textwidth]{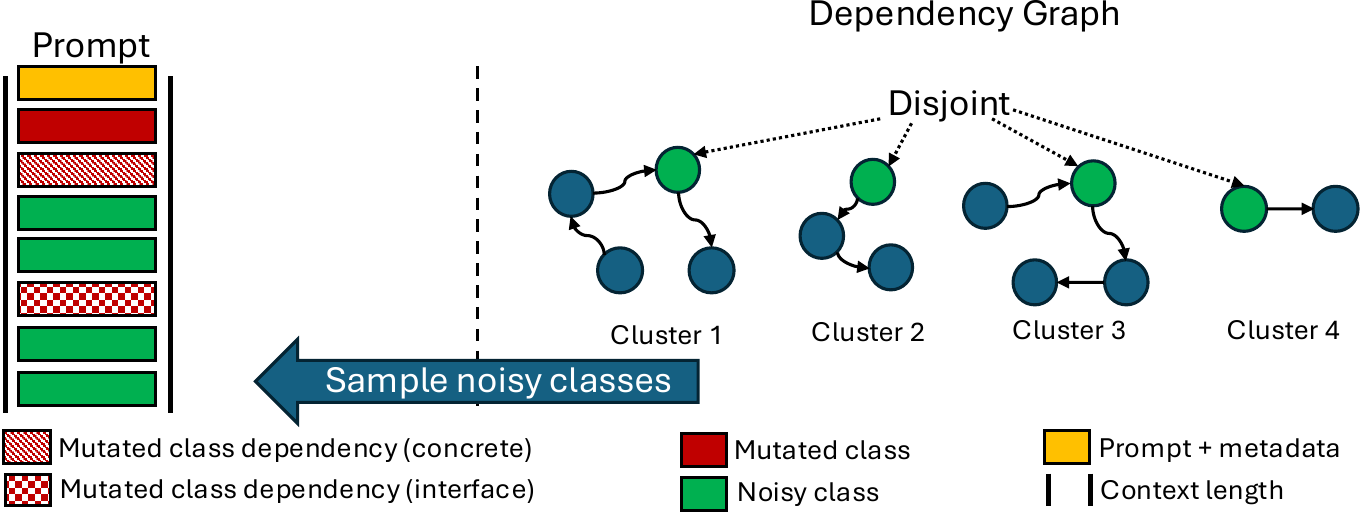}
    \caption{The selection process of \textit{noisy} classes that are used as distractor elements for the coupling task. 
    }
    \label{fig:prompt-structure}
\end{figure}

To avoid introducing any confounding effects, we carefully select such distractors. We pick disjoint classes, meaning those that are not dependent directly or \textit{transitively} on each other, or on the mutated classes. A class \textit{X} is \textit{transitively} dependent on a class \textit{Z}, if \textit{X} depends on a class \textit{Y} that depends on class \textit{Z}. To select these classes, we first generate the dependency graph of a Java project. This graph would contain \textit{dependency clusters}. Then, we generate all possible class pairs by selecting the members of disjoint clusters. Note that we also remove instantiation statements from these classes. By doing so, we fully isolate the manifestation of coupling within the classes that are mutated using the code transformations. Figure~\ref{fig:prompt-structure} illustrates how this process is conducted. 

This approach tests the model's ability to focus its analysis and discern a specific inter-class relationship amidst a field of irrelevant, but syntactically valid, code. 
To elucidate the process depicted in Figure~\ref{fig:prompt-structure}, let us consider the classes in the coupling prompt template example given in Figure~\ref{fig:templates}. In this example, \texttt{RandomNoisyClass} and \texttt{AnotherRandomNoisyClass} are two noisy classes. They do not depend on each other directly nor transitively, nor do they do so on the remaining classes. The \texttt{Service} class is the mutated class, \texttt{DepImpl} is the concrete implementation of its dependency, and \texttt{IDep} is the interface of that dependency. Under such a configuration, the only relation that exists between the classes is the tight coupling that exists between \texttt{Service} and \texttt{DepImpl}.

Together, these two methods of distortion allow us to rigorously evaluate the model's resilience and \textit{practical utility} in non-idealized conditions.

\subsection{Item Ordering}
\llm{}'s suffer from \textit{ordering bias}, where their performance is influenced by the order of items in a prompt~\cite{Xue2024, Pezeshkpour2024}. We implement a shuffling strategy to mitigate such potential bias. For cohesion analysis, which focuses on relationships within a single class, the order of the methods inside the class body is randomly shuffled. For coupling analysis, which assesses relationships between different classes, the order in which the complete class definitions appeared in the prompt, and their names depending on the task (\eg{} guided generation), is randomized. 

\subsection{Prompt Generation}
\label{sec:prompt_generation}
To generate the instruction prompts, we use the code snippets generated using the code transformations defined in Section~\ref{sbs:code_transformations} and the prompt templates mentioned in Figure~\ref{fig:templates}. However, executing all instruction prompts is prohibitively expensive. Hence, following similar works~\cite{Virk2025, Ahmed2025}, we select a subset of prompts that we use for evaluation. To ensure a rigorous and unbiased evaluation, we implement a multi-level stratified sampling technique. The complete pool of instruction prompts is first stratified by our two primary experimental factors: task type (Verification, Guided Generation, and Open-ended Generation) and distortion levels (nine levels from $0.1$/$1$ to $0.9$/$9$). Moreover, recognizing that prompt length could act as a confounding variable, we consider it as a third controlled factor. We established four same-sized bins by calculating the global quartiles ($Q_{1}$, $Q_{2}$, $Q_{3}$) of token counts across the entire population of prompts. Our final sampling frame is a three-dimensional grid defined by task type $\times$ distortion ratio $\times$ prompt length bin, where the intervals of length bins are defined as: $< Q_{1}$, $[Q1,Q2]$, $[Q_{2}, Q_{3}]$ and $> Q_{3}$. We randomly select $100$ samples from each cell in this grid to create a balanced, fully-crossed experimental design. This strategy enables a robust analysis that avoids confounding effects of task formulation, noise, and prompt length on model performance. At the end of this process, we generate $3\times9\times4\times100= 10,800$ prompts for each code transformation.
Note that the distribution of labels of the elements (classes or methods) present in the instruction portion of prompts mirrors the distortion ratio. We tie the evaluation pool so that every candidate element is scored, ensuring the model is penalized if it incorrectly labels irrelevant distractor elements.

\subsection{Models Under Test}
In our work, we focus on \textit{reasoning} models~\cite{besta2025reasoning}. Compared to standard \llm{}s that typically predict a response directly from the input, \textit{reasoning} \llm{}s distinguish themselves by generating a sequence of intermediate ``\textit{thinking tokens}'', simulating an internal Chain of Thought, before outputting the final answer. We choose this family of models given their higher performance compared to standard \llm{}s in software-related tasks~\cite{Guo2025}, and to generate the \textit{desiderata} for RQ4.

Our model selection was guided by three criteria: strong reasoning capability, scalability across model sizes, and a coverage of both open-weight and proprietary models.
First, we excluded smaller open-source reasoning models ($< 14$B) from our final evaluation after preliminary experiments revealed their frequent failure to generate responses conforming to our structured output templates. 
This hindered our ability to parse their outputs programmatically, making a systematic evaluation infeasible.
Second, to investigate the effect of scale within a reproducible open-weight category, we selected three variants from the DeepSeek-R1 model family~\cite{Guo2025}. This family is ideal for our study as it is openly available, demonstrates state-of-the-art reasoning, and offers models at different scales, enabling us to investigate the impact of model size. Furthermore, their large $128$K token context window is well-suited for processing the substantial code artifacts in our experiments. We consider that their sophisticated mathematical and logical reasoning capabilities
serve as a strong proxy for the abstract reasoning required to analyze software design properties.
Third, to broaden the coverage of our evaluation and address the generalizability of our findings to state-of-the-art proprietary models, we complement the open-weight models with two cost-effective closed-source reasoning models: GPT-$5$ Nano~\cite{Singh2025openai} and Gemini-$2.5$ Flash Lite~\cite{Comanici2025gemini}. Cost-effectiveness was an important factor as it allowed us to evaluate these models on the full datasets rather than resorting to down-sampling, thereby allowing for a fairer comparison with the set of open-weight models.

The set of models selected for evaluation summarized below.
\begin{itemize}
    \item \textbf{DeepSeek-R1 14B}: We use \textit{DeepSeek-R1-Distill-Qwen-14B}, a $14.8$B-parameter dense model created by distilling DeepSeek-R1 into Qwen$2.5$-$14$B~\cite{Qwen25}. Its strong performance for its size on benchmarks like AIME-24 (pass@1 $\approx$69.7) and MATH-500 ($\approx$93.9) demonstrates the baseline level of advanced reasoning under investigation.
    \item \textbf{DeepSeek-R1 32B}: The mid-scale variant, \textit{DeepSeek-R1-Distill-Qwen-32B} ($32.8$B parameters), follows the same distillation recipe. It achieves even higher scores on reasoning benchmarks (AIME-24 pass@1 $\approx$72.6; MATH-500 $\approx$94.3), allowing us to measure the effect of increased scale on performance.
    \item \textbf{DeepSeek-R1 70B}: For our high-end evaluation, we use \textit{DeepSeek-R1-Distill-Llama-70B} ($70.6$B parameters), which distills R1 into Llama-3.3-70B-Instruct~\cite{Grattafiori2024llama}. Its exceptional scores on complex benchmarks (e.g., GPQA-Diamond $\approx$65.2) establish it as one of the strongest openly released models for the kind of scientific and logical reasoning central to our study.

    \item \textbf{GPT-$5$ Nano}: A compact reasoning-oriented variant of OpenAI's GPT-5 family. It exposes controllable reasoning effort and supports long-context inputs (up to $400K$), making it suitable for processing the substantial code artifacts in our experiments. We include it as a representative of OpenAI's current generation of proprietary reasoning models at an accessible price point.
    \item \textbf{Gemini-$2.5$ Flash Lite}: The most cost-efficient member of Google's Gemini 2.5 family. Like the other members of the family, it is a reasoning model with a very large context window (up to $1M$). We include it as a representative of Google's current generation of proprietary reasoning models.
    
\end{itemize}
Together, these models span distinct backbone architectures (Qwen-$2.5$ and Llama-$3.3$ \added{and two proprietary ones from OpenAI and Google}), along with training regimes and a wide range of scales. This allows us to verify that our observations regarding reasoning efficacy are not artifacts of a single model family, but persist across different foundation models and parameter scales.

\subsection{Evaluation Metrics}

For verification, guided generation, and open-ended generation in the coupling task, we report the F1-score. This metric is suitable because these tasks can be framed as identifying a correct set of items: a correct binary assertion, a retrieved set of related code elements, or correctly identified pairs of coupled classes. We use the same metric for verification and guided generation for cohesion.
For open-ended generation in the cohesion task, where the goal is to \textit{partition} methods into conceptual groups based on shared responsibilities, we report the Adjusted Rand Index (ARI)~\cite{Rand1971}. ARI compares two partitions of the same set of items, in our case the methods of a class, and measures how closely the model's proposed partition aligns with the ground-truth partition. It does so by counting pairs of methods that are placed in the same group in both partitions and pairs that are placed in different groups in both partitions, and then adjusting the resulting agreement score for the level expected by chance from a random baseline. ARI is more appropriate than F1, as it evaluates the model's entire proposed partition against the ground-truth clusters, measuring their similarity while correcting for chance agreements. An ARI of $1.0$ indicates identical partitions, whereas a score near $0.0$ suggests similarity no better than random chance, and negative values indicate disagreement worse than chance from a random baseline.
These metrics are tailored to each design principle: F1 fits coupling's focus on identifying binary relationships between class pairs, while ARI addresses cohesion's inherent clustering nature by comparing proposed and ground-truth groupings.

In all experiments, we set the temperature parameter to $0$ based on the recommendations of the models' official documentation~\cite{Deepseek2025temperature}, and set the generation length to the models' maximum context size. We run inference through API calls to OpenRouter~\cite{replication}.

\subsection{Human Study}

To contextualize the performance of the \llm{}s against a human baseline, we conducted a human study in which software developers and researchers completed the same cohesion and coupling tasks presented to the models. This comparison allows us to assess whether the performance patterns observed in RQ1-RQ3 reflect limitations specific to \llms{} or are inherent to the difficulty of the tasks themselves.

\subsubsection{Instance Sampling}
Using the entire dataset, $10,800$ per concept, to human participants is infeasible. We therefore constructed a reduced, representative sample to keep the survey tractable.
We retained the three task formulations, \textit{Verification}, \textit{Guided Generation} and \textit{Open-ended Generation}, and collapsed the nine distortion levels into three coarser difficulty bins: \textit{Low} (levels $1$-$3$), \textit{Medium} (levels $4$-$6$) and \textit{Hard} (levels $7$-$9$). This reduction preserves the progression of noise while keeping the participants' workload reasonable.
To control for prompt length as a confounding factor, we restricted sampling to prompts whose length fell within the two middle bins ($[Q_{1}, Q_{2}]$ and $[Q_{2}, Q_{3}]$) defined in Section~\ref{sec:prompt_generation}. Prompts in the shortest bin ($\leq Q_{1}$) were excluded to ensure that every task offered a non-trivial amount of code to reason about, and prompts in the longest bin ($> Q_{3}$) were excluded to avoid imposing a time burden on participants. Length was treated as a filtering criterion rather than as an additional stratum since the factors of interest for the human comparison are guidance levels and distortion. From the resulting $3$ (guidance levels) $\times$ $3$ (distortion levels) we randomly sampled $7$ prompts per cell, yielding $3\times3\times7 =63$ samples per concept.

\subsubsection{Participants}

Following similar studies~\cite{Vijayvargiya2024, Feitelson2015, DiSorbo2021, Russo2019, Fucci2018}, we targeted two populations: (i) software developers with at least two years of professional experience in Java, and (ii) software engineering researchers. None of the participants was a member of the authors' list, and participation was voluntary. The survey remained open for two weeks, and participants were allowed to stop and resume later.
We were able to recruit seven participants, within the range of similar studies~\cite{Mandli2025, Zhang2026}. The participants blend academic and industry expertise. Four participants ($57.1\%$) identified themselves as academics or researchers in software engineering, two ($28.6\%$) as industry software developers, and one ($14.3\%$) as a graduate student in software engineering. In terms of Java experience, four participants ($57.1\%$) reported 2--5 years, two ($28.6\%$) reported 6--10 years, and one ($14.3\%$) reported more than $10$ years. 
Educational attainment was high: three participants ($42.9\%$) held a Master's degree, three ($42.9\%$) a Bachelor's degree, and one ($14.3\%$) a Doctoral degree.
The cohort's technical background is well-aligned with the demands of the tasks. On a four-point Likert scale measuring familiarity with object-oriented design principles, all participants rated themselves in the upper half of the scale, with four ($57.1\%$) selecting \textit{Familiar} and three ($42.9\%$) selecting \textit{Very familiar}. No participant rated themselves below \textit{Familiar}. Code review and refactoring activity was similarly concentrated at the upper end: five participants ($71.4\%$) reported performing such tasks very frequently (daily or near-daily), one ($14.3\%$) weekly, and one ($14.3\%$) occasionally. Every participant in the final sample ($100\%$) reported having previously studied or taught software design concepts in a formal setting. This indicates that the cohort had not only practical exposure to but also conceptual training in the principles being evaluated.

\subsubsection{Survey Design}
We created two separate surveys, one per design concept (cohesion and coupling), to allow participants to engage with each principle without interference from the other. Each survey presented the $63$ sampled tasks for its respective concept, with the task instruction matching the prompt template from Figure~\ref{fig:templates}. For each task, the corresponding code snippet was hosted as a GitHub Gist and linked from the survey, rather than embedded inline. We deliberately do so to let participants navigate, scroll, and inspect the code in a familiar syntax-highlighted environment, mirroring how developers read code in practice, rather than within the constrained layout of a survey form.

\subsubsection{Ground Truth and Metrics}
Participant responses are evaluated against the same ground truth used for the \llm{} evaluation, and we report the same metrics: F1-score for \textit{Verification} and \textit{Guided Generation} (both concepts) and for \textit{Open-ended Generation} in coupling, and the Adjusted Rand Index for \textit{Open-ended Generation} in cohesion. This ensures that the human and \llm{} results are directly comparable.
\section{Results}
\subsection{RQ1: Basic Understanding of Cohesion and Coupling}
Through this question, we aim to investigate the understanding of \llm{}s of each concept at a basic level. 
In Table~\ref{tab:rq1_baseline_detailed}, we present the baseline performance of the models on both coupling and cohesion under ideal, low-noise conditions. \textbf{Under such conditions, the models show a reasonable understanding of both concepts. 
However, this understanding erodes as the complexity of the task increases, suggesting a divergence in the models' practical ability to apply this knowledge.}
\begin{table}[h!]
\centering
\caption{Baseline performance ($\text{DR}_{coupling} = 0.1$, Cohesive Responsibilities = $1$) on cohesion and coupling tasks across \added{models} and transformations 
}
\label{tab:rq1_baseline_detailed}
\resizebox{\textwidth}{!}{
\begin{tabular}{lllccccc}
\toprule
\textbf{Concept} & \makecell{\textbf{Code} \textbf{Transformation}} & \textbf{\added{Cognitive Load}} & \multicolumn{5}{c}{\textbf{Model}} \\
\cmidrule(l){4-8} 
& & & \makecell{DeepSeek-R1\\($14$B)} & \makecell{DeepSeek-R1\\($32$B)} & \makecell{DeepSeek-R1\\($70$B)} &\makecell{GPT-$5$\\Nano} & \makecell{Gemini-$2.5$\\Flash Lite} \\
\midrule
\multirow{9}{*}{\textbf{Coupling}} & \multirow{3}{*}{Direct Instantiation (\did{})} & Verification & \colorcell{0.796} & \colorcell{0.885} & \colorcell{0.893} & \colorcell{0.968} & \colorcell{0.965} \\
& & Guided Generation  & \colorcell{0.607} & \colorcell{0.719} & \colorcell{0.720} & \colorcell{0.597} & \colorcell{0.693} \\
& & Open-ended Generation  & \colorcell{0.584} & \colorcell{0.629} & \colorcell{0.601} & \colorcell{0.340} &  \colorcell{0.514}\\

\cmidrule(l){2-8}
& \multirow{3}{*}{Unutilized Dependency (\uidep{})} & Verification  & \colorcell{0.810} & \colorcell{0.903} & \colorcell{0.921} & \colorcell{0.965} & \colorcell{0.969} \\ 
& & Guided Generation  & \colorcell{0.594} & \colorcell{0.731} & \colorcell{0.757} & \colorcell{0.589} & \colorcell{0.713} \\
& & Open-ended Generation  & \colorcell{0.591} & \colorcell{0.595} & \colorcell{0.609} & \colorcell{0.332} &  \colorcell{0.587} \\

\cmidrule(l){2-8}
& \multirow{3}{*}{Indirect Instantiation (\iidep{})} & Verification & \colorcell{0.758} & \colorcell{0.872} & \colorcell{0.885} & \colorcell{0.961} & \colorcell{0.954} \\
& & Guided Generation  & \colorcell{0.611} & \colorcell{0.732} & \colorcell{0.732} & \colorcell{0.564} & \colorcell{0.690} \\
& & Open-ended Generation  & \colorcell{0.591} & \colorcell{0.603} & \colorcell{0.624} & \colorcell{0.286} &  \colorcell{0.536} \\
\midrule

\multirow{3}{*}{\textbf{Cohesion}} & \multirow{3}{*}{In-cohesive Class (IC)} & Verification & \colorcell{0.799} & \colorcell{0.805} & \colorcell{0.825} & \colorcell{0.874} & \colorcell{0.770} \\
& & Guided Generation & \colorcell{0.501} & \colorcell{0.597} & \colorcell{0.649} & \colorcell{0.717} & \colorcell{0.727} \\
& & Open-ended Generation (ARI) & \colorcell{0.391} & \colorcell{0.407} & \colorcell{0.433} & \colorcell{0.461} & \colorcell{0.472} \\
\bottomrule
\end{tabular}%
}
\end{table}
\subsubsection{{Performance of Open-weight Models}}
This understanding is illustrated by the models' performance on the \textit{Verification} task, which represents the simplest form of reasoning. The $70$B model achieves an F1-score of $0.825$ for cohesion, and an average score of $0.899$ across the three coupling transformations. This pattern of comparable performance holds for the smaller  {DeepSeek-R1 variants} as well, indicating that the models' ability to recognize a well-defined instance of low cohesion or tight coupling is equally developed for both principles.

However, this balance fades when the tasks require more generative and autonomous reasoning. In the \textit{Guided Generation} level, a clear performance gap emerges in favor of coupling. The $70$B model's average F1 score for coupling ($\approx0.736$) is higher than its score for cohesion ($0.649$). This indicates that the model is more adept at identifying a discrete, related class than the task of identifying all conceptually related methods within a single class.
This performance gap widens in the fully autonomous \textit{Open-ended Generation} settings. While a direct numerical comparison between metrics is inappropriate here, we can evaluate each performance on its own terms. The model maintains a moderate F1 score for identifying coupled pairs, indicating a reasonable, albeit imperfect, retrieval capability. In contrast, its performance on cohesion is significantly challenged. The resulting ARI scores, while better than random, are far from indicating a coherent clustering. This highlights the challenge the model faces when it must autonomously discover multiple, distinct responsibilities within a class.

In addition, two further observations can be drawn. First, a consistent trend across all conditions is the positive impact of the size of the model. In nearly every experiment for both concepts, performance improves as model size increases from $14$B to $70$B, demonstrating that a more sophisticated grasp of abstract software design is an emergent property of scale {within this family}. Second, the models exhibit some robustness to implementation details in the coupling tasks. The performance remains relatively close across the \did{}, \uidep{} and \iidep{} transformations, suggesting that the models have developed a generalized, conceptual understanding of dependency in a low-noise environment.

\subsubsection{{Performance of Proprietary Models}}

{We now turn to the two proprietary models, GPT-$5$ Nano and Gemini-$2.5$ Flash Lite. As their parameter counts are undisclosed, we refrain from attributing the differences between them to scale and report their results as a separate category.}  

{On the \textit{Verification} task, both models achieve strong performance on coupling, with average F1 scores of $\approx0.965$ for GPT-$5$ Nano and $\approx0.963$ for Gemini-$2.5$ Flash Lite. On cohesion, however, they diverge: GPT-$5$ Nano achieves $0.874$ while Gemini-$2.5$ Flash Lite reaches only $0.770$. \textit{Verification} performance within the proprietary category is therefore concept-dependent: the two models perform comparably on coupling but diverge on cohesion.}

{On \textit{Guided Generation}, the two models exhibit contrasting behavior on coupling. GPT-$5$ Nano's average F1 drops to $\approx0.583$, whereas Gemini-$2.5$ Flash Lite holds at $\approx0.699$. For cohesion, both perform closely, with F1 scores of $0.717$ and $0.727$, respectively. Notably, both models score higher on cohesion than on coupling under guided generation, reversing the ordering observed in verification.}

{On \textit{Open-ended Generation}, they separate further on coupling. GPT-$5$ Nano's F1 falls to $\approx0.319$, while Gemini-$2.5$ Flash Lite retains a higher score of $\approx0.546$. For cohesion, both models produce comparable ARI scores ($0.461$ and $0.472$). Both also remain stable across the three coupling transformations, suggesting that their recognition of dependency is not tied to a specific syntactic manifestation. Within the proprietary category, the two models differ mainly in coupling, and this gap widens as cognitive load increases.}

\subsubsection{{Cross-category Observations}}

{When comparing both categories within this idealized, low-noise setting, three observations emerge. First, on the \textit{Verification} task for coupling, both proprietary models ($\approx0.96$) score higher than the strongest open-weight model ($\approx0.900$ for the $70$B). For cohesion verification, the results are mixed: GPT-$5$ Nano ($0.874$) exceeds the $70$B open-weight model ($0.825$), while Gemini-$2.5$ Flash Lite ($0.770$) trails even the $14$B open-weight variant ($0.799$).}

{Second, the proprietary models' lead on verification does not extend to the more autonomous settings on coupling. Under \textit{Guided Generation}, GPT-$5$ Nano's coupling F1 ($\approx0.583$) drops below the $14$B open-weight variant ($\approx0.604$), while Gemini-$2.5$ Flash Lite ($\approx0.699$) remains close to but below the $70$B open-weight model ($\approx0.736$). Under \textit{Open-ended Generation}, GPT-$5$ Nano ($\approx0.319$) falls below every open-weight model, while Gemini-$2.5$ Flash Lite ($\approx0.546$) sits between the $14$B and $70$B. The Verification-to-Open-ended drop for GPT-$5$ Nano on coupling ($\sim67\%$) is considerably steeper than the corresponding drop for the $70$B open-weight model ($\sim$$32\%$). For cohesion, by contrast, the proprietary models match or slightly exceed the open-weight models across all three cognitive loads, and the two proprietary models yield the highest ARI scores in the open-ended partitioning task ($0.461$ and $0.472$, vs. $0.433$ for the $70$B).}

{Third, the performance is stable across the \did{}, \uidep{}, and \iidep{} transformations within both categories. This indicates that recognition of dependency is not tied to a specific manifestation of tight coupling across either category. Taken together, these observations suggest that category-level differences are much more pronounced for coupling than for cohesion, and that strong \textit{Verification} performance, which was achieved by GPT-$5$ Nano, does not translate into robustness under \textit{Open-ended Generation}.}

\begin{boxH}
 \textbf{Summary}: While the {open-weight and proprietary models} demonstrate a reasonable grasp when verifying cohesion and coupling in a low-noise environment, their ability to apply this knowledge diverges as tasks demand more autonomy. {The proprietary models lead on \textit{Verification} for coupling but exhibit the steepest drop under \textit{Open-ended Generation}, whereas on cohesion the two categories perform comparably.}
\end{boxH}

\subsection{RQ2: Impact of Cognitive Load on Model Performance}
The level of contextual help provided in a prompt is a dominant factor influencing model performance. \textbf{For both cohesion and coupling, model performance is highest in \textit{Verification} tasks, which provide maximum guidance, followed by \textit{Guided Generation} tasks with partial guidance, and is lowest in \textit{Open-ended Generation} tasks that offer minimal help}.
\subsubsection{{Performance of Open-weight Models}}
\begin{figure}[H]
\includegraphics[width=\textwidth]{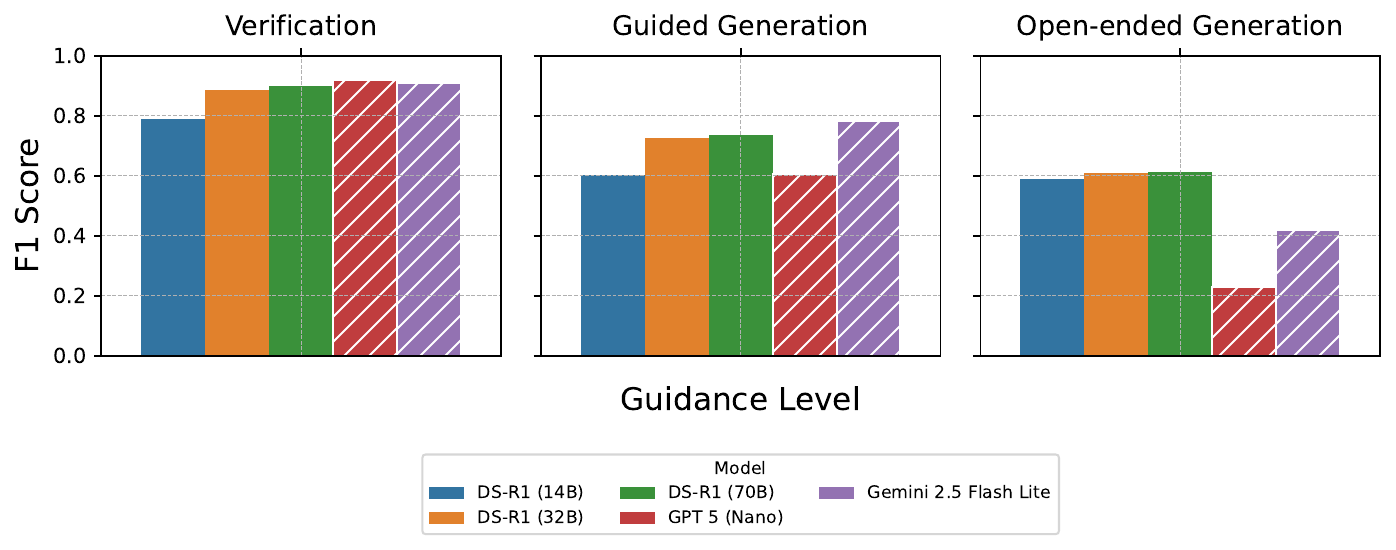}
\caption{Impact of the cognitive load on the performance of each model in coupling analysis.}
\label{fig:rq2_coupling}
\end{figure}

For coupling tasks, performance declines as guidance is reduced. This trend holds across all model scales, though the nature of the decline varies with model size. For instance, the $70$B model's score falls from a high of $0.90$ in \textit{Verification} to $0.74$ in \textit{Guided generation}, and finally to $0.61$ in the Open-ended setting, representing a performance decrease of over $32\%$ from its peak. Note that the figures represent the average across the three code transformations.

Larger models demonstrate more resilience when task complexity initially increases. The $32$B and $70$B models show a smaller performance drop of approximately $18\%$ when moving from \textit{Verification} to \textit{Guided generation}, compared to $23\%$ for the $14$B model. However, this advantage diminishes when faced with the complexity of open-ended generation. The subsequent performance drop from \textit{Guided} to \textit{Open-ended} is far more severe for the larger models ($\sim$$17\%$) than for the $14$B model ($\sim$$2.5\%$).

\begin{figure}[h!]
\centering
\includegraphics[width=\linewidth]{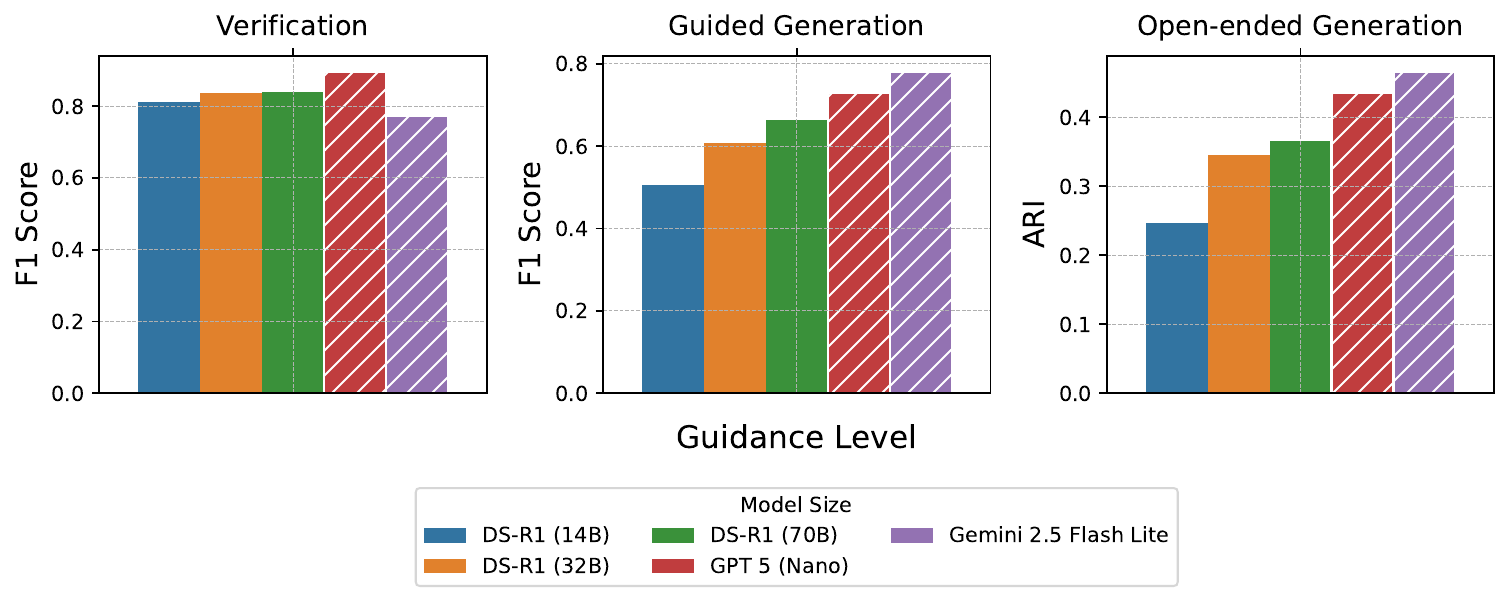}
\caption{Impact of the cognitive load on the performance of each language model on the cohesion experiments.}
\label{fig:rq2_cohesion}
\end{figure}

For cohesion, reducing guidance also lowers performance, as illustrated in Figure~\ref{fig:rq2_cohesion}. {\textit{Verification} and \textit{Guided Generation} are scored with F$1$, while \textit{Open-ended Generation} is scored with ARI, since recovering the full partition is a clustering problem.} In the \textit{Verification} task, all models perform strongly, achieving F$1$-scores that range between $0.79$ and $0.82$. {Performance is lower under \textit{Guided Generation}, which asks the model to identify all methods related to a given \textit{seed} method. For instance, the $70$B model's F$1$-score reaches around $0.66$. On the \textit{Open-ended Generation} task, the top ARI score among the open-weight models is approximately $0.37$. F$1$ and ARI are on different scales and are not directly comparable, but together they indicate that while the model is proficient at validating a specific cohesive relationship, autonomously discovering all of a class's underlying responsibilities is significantly more challenging.}

\subsubsection{{Performance of Proprietary Models}}
{For coupling, the performance of both models declines as guidance is reduced, but GPT-$5$ Nano's falls more steeply than Gemini-$2.5$ Flash Lite's. GPT-$5$ Nano drops from $0.918$ in \textit{Verification} to $0.605$ in \textit{Guided Generation}, and then to $0.230$ in \textit{Open-ended Generation}, which represents a $\sim$$75\%$ drop from its peak performance. On the other hand, Gemini-$2.5$ Flash Lite declines more gradually, from $0.910$ to $0.783$ and then to $0.421$. This represents a drop of $\sim$$54\%$. Both models start from a similar \textit{Verification} score but separate as the task requires greater autonomy.}

{For cohesion, the two models respond to reduced guidance differently. GPT-$5$ Nano's F$1$ scores are $0.896$ on \textit{Verification} and $0.729$ on \textit{Guided Generation}, and its ARI on \textit{Open-ended Generation} is $0.434$. Gemini-$2.5$ Flash Lite's F$1$ scores under \textit{Verification} ($0.772$) and \textit{Guided Generation} ($0.781$) are close to one another, and its ARI on \textit{Open-ended Generation} is $0.466$. Although the F$1$ measure and ARI are not directly comparable, the recurring pattern is that GPT-$5$ Nano's lead under guidance does not carry over to the autonomous setting, while Gemini's performance stays roughly flat between \textit{Verification} and \textit{Guided Generation} before degrading once guidance is removed.}

\subsubsection{{Cross-category Observations}}
{In general, reducing the guidance lowers the performance of every model. The exception is Gemini's, where the difference between its \textit{Verification} and \textit{Guided Generation} performance is narrow, almost identical ($\Delta F1 = 0.8$). What differs between categories in the size of each drop and the ordering of the models at each cognitive level.}

{For coupling, the proprietary models lead on \textit{Verification} but lose that lead once autonomy is required. At \textit{Open-ended Generation}, GPT-$5$ Nano and Gemini-$2.5$ Flash Lite achieve F1 scores of $0.230$ and $0.421$, respectively. Such performance trails behind all open-weight models whose performance, at such a cognitive level, is between $0.589$ and $0.611$. The Verification$\rightarrow$Open-ended Generation drop rates for GPT-$5$ Nano ($\sim$$75\%$) and for Gemini ($\sim$$54\%$), both exceed that of the $70$B open-weight model ($\sim$$32\%$). Strong verification on coupling, therefore, does not entail robustness when the model must discover dependencies without guidance.}

{As for cohesion, ordering between families is reversed as guidance is reduced. On \textit{Verification}, the $70$B open-weight model ($0.841$) outperforms Gemini-$2.5$ Flash Lite ($0.772$) but trails GPT-$5$ Nano ($0.896$). On \textit{Guided Generation}, both proprietary models, Gemini at $0.781$ and Nano at $0.729$, exceed DeepSeek-R$1$ ($70$B)'s performance of $0.665$. The same trend is observed on \textit{Open-ended Generation}, where both lead with ARI scores of $0.466$ and $0.434$ versus $0.365$ for the $70$B.}
\begin{boxH}
\textbf{Summary}: 
For both concepts, performance drops significantly as tasks shift from verification to open-ended generation.{This pattern holds across both open-weight and proprietary model families.} Models excel at verifying principles but struggle when {faced with scenarios that require autonomy}. {For coupling, the drop is steepest for GPT-$5$ Nano, whose F1 falls from $0.918$ in \textit{Verification} to $0.230$ in \textit{Open-ended Generation}, a decrease that exceeds the corresponding $\sim$$32\%$ drop for the $70$B open-weight model. For cohesion, no model exceeds an ARI of $0.47$ on the open-ended task.}

\end{boxH}

\subsection{RQ3: Robustness Against Distortion}
To investigate the impact of increasing distortion on model performance for cohesion and coupling, we analyze the 
evaluation metrics as the controlled noise level is increased. \textbf{Our findings reveal a significant difference in robustness, with the models' understanding of cohesion being more resilient to noise than their understanding of coupling.}

\begin{figure}[H]
    \centering
    \includegraphics[width=\linewidth]{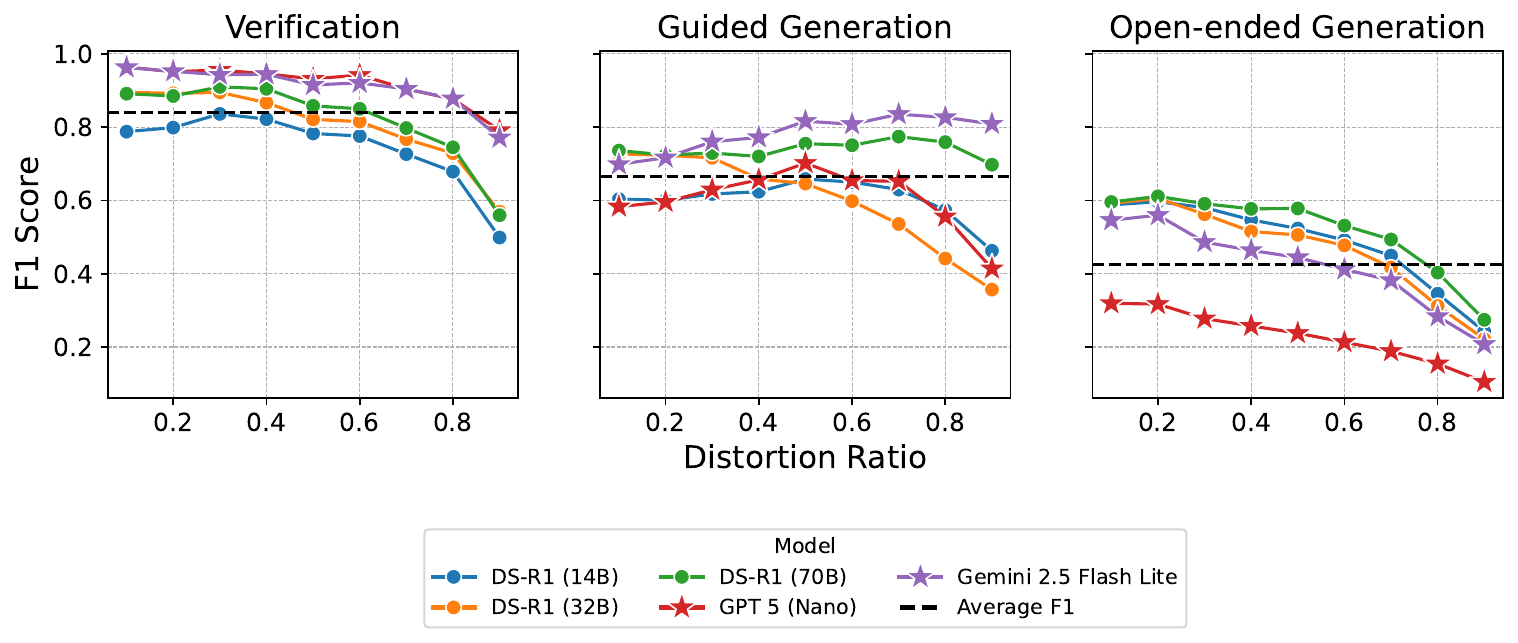}
    \caption{Performance of {all models} on the coupling task under each distortion ratio. Results are aggregated across all code transformations. The average F1 score refers to the mean across models and distortion levels.}
    \label{fig:coupling_results_figures}
\end{figure}

The results for the coupling tasks, as shown in Figure~\ref{fig:coupling_results_figures}, demonstrate a consistent decline in performance across all models and {cognitive loads} as the distortion ratio increases. 

\subsubsection{{Performance of Open-weight Models}}

The results indicate that the models' ability to reason about inter-module dependencies is highly sensitive to the presence of unrelated \textit{distractor} classes. In the most challenging task, \textit{Open-ended Generation}, performance degrades sharply and steadily for all models. For instance, the F1 score for the $70$B model drops from $0.59$ at a low distortion ratio of $0.1$ to just $0.27$ at a high ratio of $0.9$, representing a performance decrease of over $54$\%. A similar trend is observed for the $14$B and $32$B models, whose scores fall to $0.24$ and $0.22$, respectively, confirming that even larger models struggle with autonomous discovery in a noisy environment.

A similar, albeit less severe, degradation is observed for the \textit{Guided Generation} task. While the $70$B model exhibits stability, maintaining an F1 score above $0.70$ until the highest distortion levels, the smaller models are more fragile. The $32$B model's performance, for example, erodes significantly from a peak of $0.72$ down to $0.35$ as the distortion ratio rises from $0.1$ to $0.9$. Even in the \textit{Verification} task, which provides the most contextual help, performance is not immune to noise. All models maintain high F1 scores (often above $0.80$) for distortion ratios up to $0.6$, but their performance begins to decline noticeably thereafter. For example, the $14$B model's F1 score falls from a peak of $0.83$ to below $0.50$ at the highest distortion level. This suggests that as the context becomes saturated with irrelevant classes, the models' ability to confirm a specific dependency is affected. Overall, the consistent performance drop across all coupling tasks indicates that reasoning about inter-module relationships is vulnerable to contextual noise.

\begin{figure}[H]
    \centering
    \includegraphics[width=\linewidth]{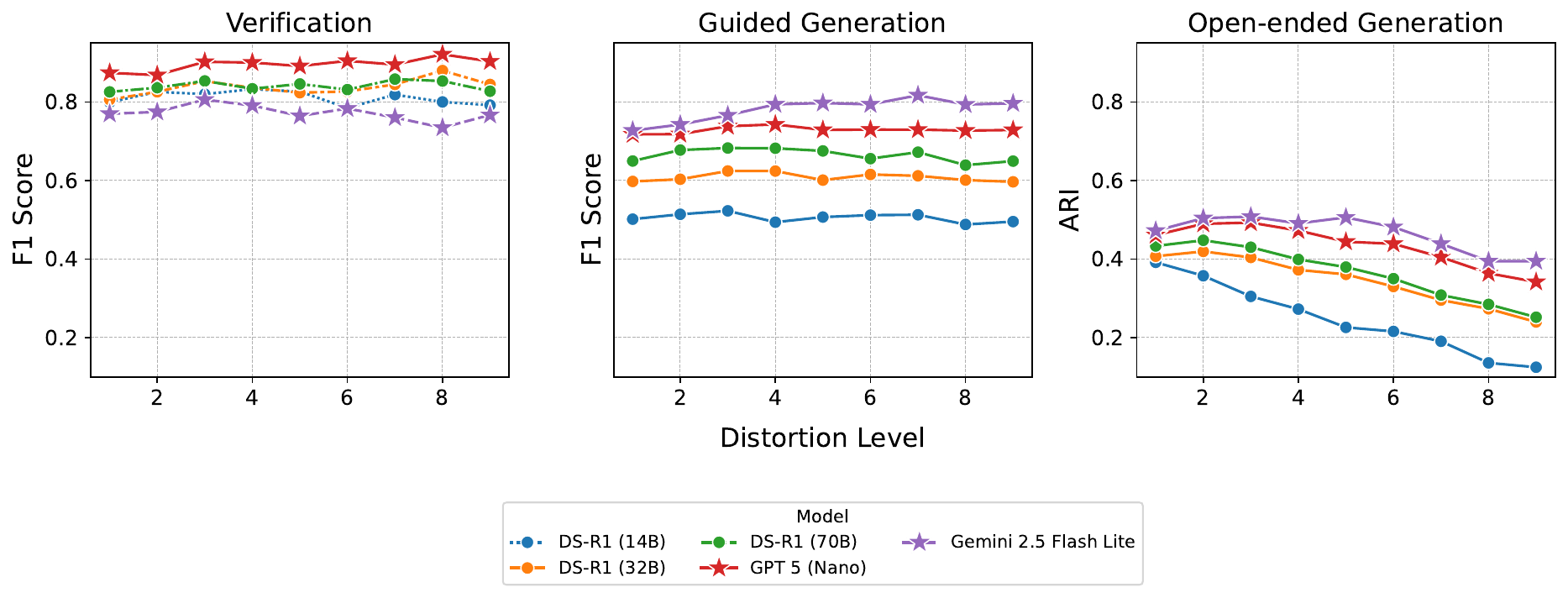}
    \caption{Performance of {all models} on the cohesion task under each distortion level.}
    \label{fig:cohesion_results_figures}
\end{figure}

{In Guided Generation, the F1 scores form an inverted-U shape as the share of tightly coupled pairs drops from $\sim$$90\%$ (at $0.1$ distortion) to $\sim$$10\%$ (at $0.9$). The $70$B model remains stable because its high precision offsets lost recall, whereas the $14$B and $32$B models degrade sooner by over-predicting positives.}

In contrast, the models exhibit robustness when performing cohesion tasks, particularly for the \textit{Verification} and \textit{Guided Generation} prompts. The performance on these tasks remains remarkably stable across all distortion levels, from $1$ to $9$. For the \textit{Verification} task, the F1 scores for all three models consistently hover between $0.80$ and $0.85$, with no discernible negative trend as the number of injected source classes increases. For instance, the $70$B model achieves an F1 score of $0.825$ at a distortion level of 1 and a score of $0.827$ at a distortion level of 9, showing no degradation. This implies that the models' ability to verify the cohesiveness of a code block is not affected by the degree of conceptual pollution within the class.

A similar pattern of resilience is illustrated in the \textit{Guided Generation} task. The F1 scores for all models remain almost flat across the entire range of distortion levels. The $32$B model, for example, starts with an F1 score of $0.597$ at a distortion level of $1$ and ends with a score of $0.596$ at a level of $9$. This stability suggests that when given a \textit{seed} method, the models can effectively identify its related peers by focusing on local intra-method relationships (like shared field access) while successfully ignoring the noise from conceptually unrelated methods.

The sole exception to this pattern of robustness is the \textit{Open-ended Generation} task for cohesion. In this scenario, where models must partition all methods into responsibility clusters without any guidance, performance, as measured by the Adjusted Rand Index, declines steadily with increasing distortion. The ARI for the $70$B model, for instance, decreases from a peak of $0.44$ at a distortion level of 2 to $0.25$ at a level of $9$. This shows that while models are robust at analyzing local cohesion, their ability to form a global, correct partitioning of a class's responsibilities is \textit{fragile} and degrades as conceptual noise increases.

\subsubsection{{Performance of Proprietary Models}}
{For coupling, both proprietary models decline on \textit{Open-ended Generation} as distortion increases. GPT-$5$ Nano falls from $0.319$ F1 score at a distortion ratio of $0.1$ to $0.105$ F1 score at $0.9$, a $\sim$67\% drop. Gemini-$2.5$ Flash Lite falls from $0.546$ to $0.208$ over the same range, a $\sim$$62\%$ drop.}

{The declines on \textit{Verification} and \textit{Guided Generation} for coupling are smaller, and the two models behave differently on \textit{Guided Generation}. On \textit{Verification}, GPT-$5$ Nano's F1 falls by $\sim$$18\%$, going from $0.964$ at a distortion ratio of $0.1$ to $0.791$ at $0.9$. As for Gemini's performance, it exhibits a drop of $\sim$$20\%$, falling from from $0.963$ to $0.771$. On the \textit{Guided Generation}, GPT-$5$ Nano peaks at $0.702$ at a distortion ratio of $0.5$ before dropping to $0.414$ at $0.9$. Gemini, by contrast, rises almost monotonically from $0.669$ at $0.1$ to $0.809$ at $0.9$, peaking at $0.836$ at $0.7$.}

{For cohesion, the two models remain stable on \textit{Verification}, while on \textit{Guided Generation} only GPT-$5$ Nano is stable and Gemini rises with distortion. On \textit{Verification}, GPT-$5$ Nano's F1 statys between $0.868$ and $0.921$, and Gemini's between $0.735$ and $0.817$ with no monotonic trend in either case. On \textit{Guided Generation}, GPT-$5$ Nano's performance remains between $0.717$ and $0.743$. Gemini's rises from $0.727$ at level $1$ to $0.796$, peaking at $0.817$ at level $7$, mirroring the upward trend we observed on coupling.}

{Finally, on the \textit{Open-ended Generation} task for cohesion, the performance of both models declines as distortion increases. GPT-$5$ Nano's ARI falls from $0.461$ at level $1$ to $0.342$ at level $9$. Gemini's ARI falls from $0.472$ to $0.395$.}

\subsubsection{{Cross-category Observations}}

{The impact of distortion on each concept depends on the cognitive level. On coupling, performance declines at nearly every cognitive level as distortion increases. All five models decline on \textit{Verification} and \textit{Open-ended Generation}, and four of the five also decline on \textit{Guided Generation}. On cohesion only \textit{Open-ended Generation} shows a decline, whereas \textit{Verification} and \textit{Guided Generation} remain relatively stable across the distortion range for every model, with Gemini's rising \textit{Guided Generation} trend as the only exception. This general pattern holds across both categories. A plausible interpretation is that reasoning about inter-module dependencies requires a global view of the codebase that is easily disrupted by irrelevant classes, especially under settings where exploration is needed (\ie{} \textit{Open-ended Generation}). On the other hand, intra-class cohesion can be assessed from local signals such as shared field access and direct method-to-method interaction, which remain largely unaffected by injected distractor methods.}

{The two categories, open-weight and proprietary, differ in how quickly the performance on the \textit{Verification} level degrades. At a distortion ratio of $0.9$, the open-weight models fall to F1 between $0.499$ for the DeepSeek-R1 ($14$B) and $0.569$ for $32$B variant. On the other hand, GPT-$5$ Nano and Gemini-$2.5$ Flash Lite reach F1 scores of $0.791$ and $0.771$, respectively. This entails that under such conditions, the closed-source models have stronger filtering capabilities, allowing them to retain context better.}

{However, this advantage on the coupling task does not extend to \textit{Open-ended Generation}. Both proprietary models record lower scores than those of the open-weight models across all distortion levels.}
{The rising trend in Gemini's performance as distractor elements are added under the \textit{Guided Generation} guidance can be explained following the same logic previously discussed in the open-weight results interpretation. On coupling, its F1 moves from $0.699$ at a distortion ratio of $0.1$ to $0.809$ at $0.9$, peaking at $0.836$ at $0.7$. On cohesion, its F1 moves from $0.727$ at level $1$ to $0.796$ at level~$9$, peaking at $0.817$ at level~$7$.}

{On \textit{Open-ended Generation} for cohesion, the proprietary models sustain higher ARI scores throughout the distortion range than any open-weight variant. At distortion level $9$, GPT-$5$ Nano reaches $0.342$ and Gemini $0.395$, compared to $0.252$ for the $70$B open-weight model. The proprietary models, therefore, provide a stronger baseline for unguided responsibility partitioning.}

\begin{boxH}
    \textbf{Summary}: Our findings highlight that reasoning about software coupling is highly sensitive to noise, with performance declining steadily as distractor classes are introduced, whereas reasoning about cohesion is more robust, and the sensitivity is rather impacted by the task complexity. They also illustrate that inter-class reasoning is more challenging in noisier environments compared to intra-class reasoning. {This behavior holds across both open-weight and proprietary categories. The proprietary models retain a larger share of their coupling \textit{Verification} accuracy under high distortion and sustain higher ARI scores on open-ended cohesion, though coupling remains the more fragile concept for every model we evaluated.}
\end{boxH}
{\subsection{RQ4: Human Study}}
{To contextualize the performance of the \llm{}s against a human baseline, we conducted a human study where human participants completed the same cohesion and coupling tasks. \textbf{The main finding is that LLMs operate at the competence level of an ``intern''; they perform respectably on well-structured tasks with explicit cues, but the human-\llm{} performance gap widens as tasks demand more autonomy and noise increases.}}
{We exclude tasks that were answered by fewer than two human participants~\cite{Vijayvargiya2024}. Hence, we are unable to report the results of \textit{Verification} under the low distortion ratio for coupling, and the results of \textit{Guided Generation} under high distortion for cohesion. Nonetheless, we observe a consistent pattern across the remaining settings.}
\begin{table}[ht]
\centering
\caption{Human vs. \llm{} performance on coupling tasks. Each cell reports the mean F1 across tasks for the human cohort and each evaluated model. Per-task human scores are aggregated as the mean across responding participants. Tasks answered by fewer than two participants are excluded. Cell sample sizes are small (at most $n{=}7$ paired tasks per cell), and we therefore report only descriptive statistics. The best performance is in \textbf{bold}, whereas the highest \llm{} performance is \underline{underlined}.}
\label{tab:human-vs-llm-coupling}
\resizebox{\textwidth}{!}{
\begin{tabular}{llrrrrrr}
\toprule
Cognitive Load & Noise Level & Humans & DeepSeek-R1-14B & DeepSeek-R1-32B & DeepSeek-R1-70B & GPT 5 (Nano) & Gemini 2.5 Flash Lite \\
 &  & $\overline{F1}$ & $\overline{F1}$ & $\overline{F1}$ & $\overline{F1}$ & $\overline{F1}$ & $\overline{F1}$ \\
\midrule
Verification & Medium & \textbf{0.998} & 0.966 & 0.971 & \underline{0.999} & \underline{0.999} & 0.992 \\
Verification & Hard & \textbf{0.997} & 0.905 & 0.911 & 0.983 & \underline{0.989} & 0.988 \\ 
\midrule
Guided Generation & Low & \textbf{0.923} & 0.731 & 0.720 & \underline{0.744} & 0.722 & 0.739 \\
Guided Generation & Medium & \textbf{0.871} & 0.692 & 0.701 & 0.725 & 0.691 & \underline{0.757} \\
Guided Generation & Hard & \textbf{0.823} & 0.681 & 0.690 & \underline{0.733} & 0.551 & 0.721 \\
\midrule
Open-ended Generation & Low & \textbf{0.778} & 0.512 & 0.578 & \underline{0.691} & 0.444 & 0.492 \\
Open-ended Generation & Medium & \textbf{0.669} & 0.560 & 0.614 & \underline{0.633} & 0.407 & 0.487 \\
Open-ended Generation & Hard & \textbf{0.483} & 0.301 & 0.307 & \underline{0.377} & 0.158 & 0.239 \\
\bottomrule
\end{tabular}%
}
\end{table}

{The performance gap is narrowest in the most structured task formulation. In the \textit{Verification} tasks, the prompt enumerates the code elements, and the model only needs to confirm a stated relationship. Under these intern-friendly conditions, \llm{}s perform at a respectable level. For coupling \textit{Verification}, top \llm{}s score between $0.983$ and $0.999$, essentially tying with the human cohort's $0.997$ to $0.998$, as shown in Table~\ref{tab:human-vs-llm-coupling}. For cohesion \textit{Verification}, humans score roughly $0.950$, while top \llm{} achieve between $0.786$ and $0.882$, as indicated in Table~\ref{tab:human-vs-llm-cohesion-main}. This shows a real but small gap where the models effectively utilize the provided scaffolding to succeed.}
\begin{table}[ht]
    \centering
    \caption{Human vs. \llm{} performance on cohesion tasks. Each cell reports the mean F1 and ARI across tasks for the human cohort and each evaluated model. Per-task human scores are aggregated as the mean across responding participants. Tasks answered by fewer than two participants are excluded. Cell sample sizes are small (at most $n{=}7$ paired tasks per cell), and we therefore report only descriptive statistics. The best performance is in \textbf{bold}, whereas the highest \llm{} performance is \underline{underlined}.}
    \label{tab:human-vs-llm-cohesion-main}

    \begin{subtable}[t]{\textwidth}
        \centering
        \caption{Results for the \textit{Verification} and \textit{Guided Generation} cognitive loads.}
        \label{tab:sub-f1}
        \resizebox{\linewidth}{!}{
            \begin{tabular}{llrrrrrr}
            \toprule
            Cognitive Load & Noise Level & Humans & DeepSeek-R1-14B & DeepSeek-R1-32B & DeepSeek-R1-70B & GPT 5 (Nano) & Gemini 2.5 Flash Lite \\
            & & $\overline{F1}$ & $\overline{F1}$ & $\overline{F1}$ & $\overline{F1}$ & $\overline{F1}$ & $\overline{F1}$ \\
            \midrule
            Verification & Low & \textbf{0.962} & 0.770 & 0.779 & 0.815 & \underline{0.882} & 0.792 \\
            Verification & Medium & \textbf{0.950} & 0.759 & 0.770 & 0.802 & \underline{0.832} & 0.786 \\
            Verification & Hard & \textbf{0.949} & 0.750 & 0.771 & 0.789 & \underline{0.801} & 0.770 \\
            \midrule
            Guided Generation & Low & \textbf{0.882} & 0.550 & 0.736 & 0.746 & \underline{0.765} & 0.761 \\
            Guided Generation & Medium & \textbf{0.870} & 0.533 & 0.724 & 0.733 & \underline{0.759} & 0.750 \\
            \bottomrule
            \end{tabular}%
        }
    \end{subtable}
    
    \begin{subtable}[t]{\textwidth}
        \centering
        \caption{Results for the \textit{Open-ended Generation} cognitive load.}
        \label{tab:sub-ari}
        \resizebox{\linewidth}{!}{
            \begin{tabular}{llrrrrrr}
            \toprule
            Cognitive Load & Noise Level & Humans & DeepSeek-R1-14B & DeepSeek-R1-32B & DeepSeek-R1-70B & GPT 5 (Nano) & Gemini 2.5 Flash Lite \\
            & & $\overline{ARI}$ & $\overline{ARI}$ & $\overline{ARI}$ & $\overline{ARI}$ & $\overline{ARI}$ & $\overline{ARI}$ \\
            \midrule
            Open-ended Generation & Low & \textbf{0.791} & 0.520 & 0.559 & 0.581 & 0.614 & \underline{0.620} \\
            Open-ended Generation & Medium & \textbf{0.780} & 0.461 & 0.485 & 0.493 & 0.511 & \underline{0.534} \\
            Open-ended Generation & Hard & \textbf{0.704} & 0.290 & 0.312 & 0.369 & 0.408 & \underline{0.419} \\
            \bottomrule
            \end{tabular}%
        }
    \end{subtable}
\end{table}

{The human-\llm{} performance gap grows as guidance is reduced. In \textit{Guided Generation}, the best \llm{} trails humans by approximately $15$ to $19$ points on coupling and $11$ to $14$ points on cohesion. The gap widens even further in \textit{Open-ended Generation} tasks. While \llm{}s hold up well under simple recognition tasks, they struggle as guidance is reduced.}

{Contextual noise disproportionately affects \llm{}s compared to humans. In \textit{Verification} tasks, both humans and \llm{}s maintain relatively high performance across medium and hard distortion levels. However, in \textit{Open-ended} tasks, human performance degrades gracefully. From low to hard noise levels, human performance drops by about $38\%$ in coupling and $11\%$ in cohesion, computed as $(\text{score}_{\text{low}} - \text{score}_{\text{hard}}) / \text{score}_{\text{low}}$. In contrast, \llm{} performance drops by $32\%$ to $64\%$ across the same conditions. Reducing guidance while simultaneously injecting noise is where human participants excelled.}

{The most challenging experimental setting reveals the largest capability gap. In this setting, humans hold their largest leads, scoring $0.483$ versus the best \llm{}'s $0.377$ on coupling, and $0.704$ versus the best \llm{}'s $0.419$ on cohesion. Furthermore, humans achieve top performance across both design concepts uniformly, unlike \llm{}s. For instance, GPT-5 Nano is the strongest model on cohesion \textit{Verification} but the weakest on coupling \textit{Open-ended Generation}. This lack of broad, uniform competence further underscores their intern-like profile.}

\begin{boxH}
\textbf{Summary}: {Our comparison with a human baseline confirms the intern-level profile of some of the current language models. While \llm{}s match human accuracy on highly structured verification tasks, they fall behind when forced to autonomously discover and evaluate design flaws in noisy environments. Furthermore, unlike human developers who exhibit robust and uniform competence across different software design concepts, the models show inconsistent performance.}
\end{boxH}

\subsection{RQ{5}: Trends in Reasoning Traces}
\label{section:trends_in_reasoning_traces}
We investigated the trends in the reasoning traces generated by the \llm{}s. \textbf{When {open-weight models} analyze classes for coupling, they generate unexpectedly shorter traces on the open-ended generation task compared to those of verification and guided generation. Further experiments reveal {that this category} was incapable of analyzing all the classes that they were presented with. As for cohesion, we observed that the increase in cognitive load leads to longer traces.}

\subsubsection{Trends during coupling analysis}
\begin{figure}[H]
     \centering
     \begin{subfigure}[b]{\textwidth}
         \centering
         \includegraphics[width=\textwidth]{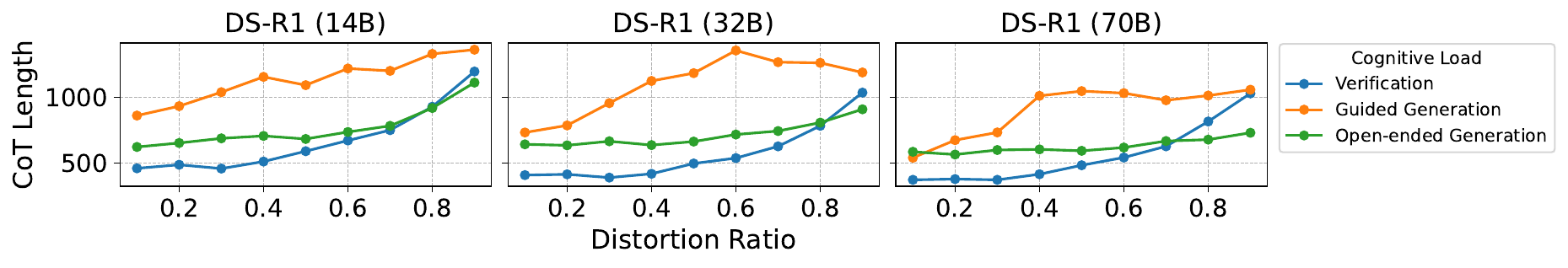}
         \caption{{Open-weight models}}
         \label{fig:rq5_coupling_openweight}
     \end{subfigure}
     
     \vspace{1em} 

     \begin{subfigure}[b]{.7\textwidth}
         \centering
         \includegraphics[width=\textwidth]{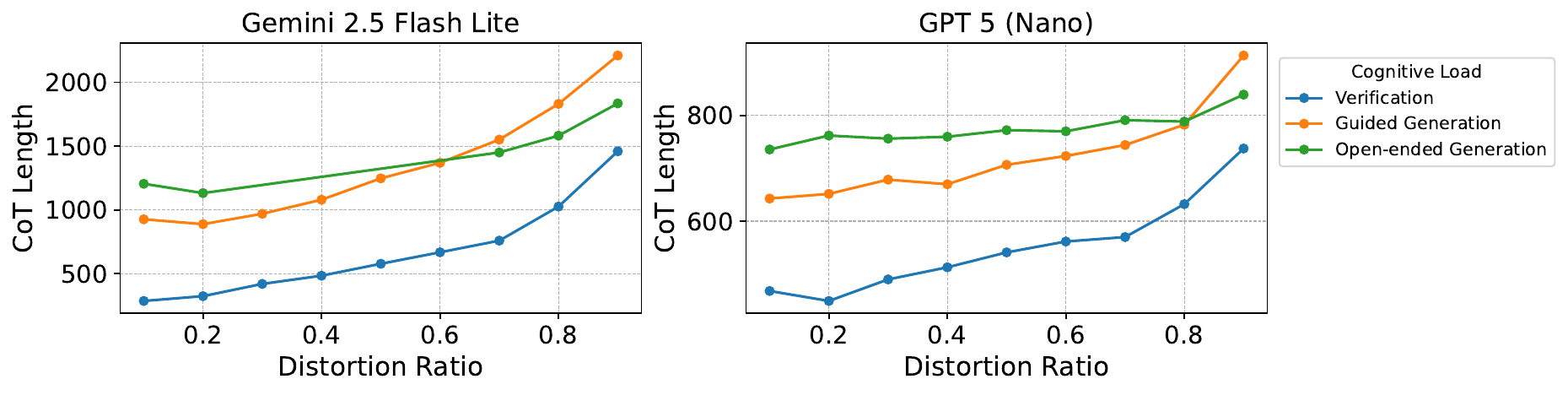}
         \caption{{Proprietary models}}
         \label{fig:rq5_coupling_prop}
     \end{subfigure}
     \caption{Relationship between the distortion ratio and the token count of reasoning traces across all models on the coupling task. We use the \did{} code transformation results, given that the coupling code transformations do not impact the models' performance.}
     \label{fig:cot_len_vs_dr_coupling}
\end{figure}

As illustrated in Figure~\ref{fig:rq5_coupling_openweight}, the average chain-of-thought (CoT)
length (in tokens) does not directly correlate with the presumed difficulty of the task. The simplest task, \textit{Verification}, consistently produces the shortest traces. However, contrary to expectations, the \textit{Open-ended Generation} task, which demonstrated the lowest performance in our prior analysis and requires the most independent discovery, generates shorter reasoning traces than the \textit{Guided Generation} task. This finding suggests that a higher cognitive load does not necessarily lead to a more exhaustive deliberation process when {open-weight models} reason about coupling.

\begin{figure}[!t]
     \centering
     \begin{subfigure}[b]{\textwidth} 
         \centering
         \includegraphics[width=\textwidth]{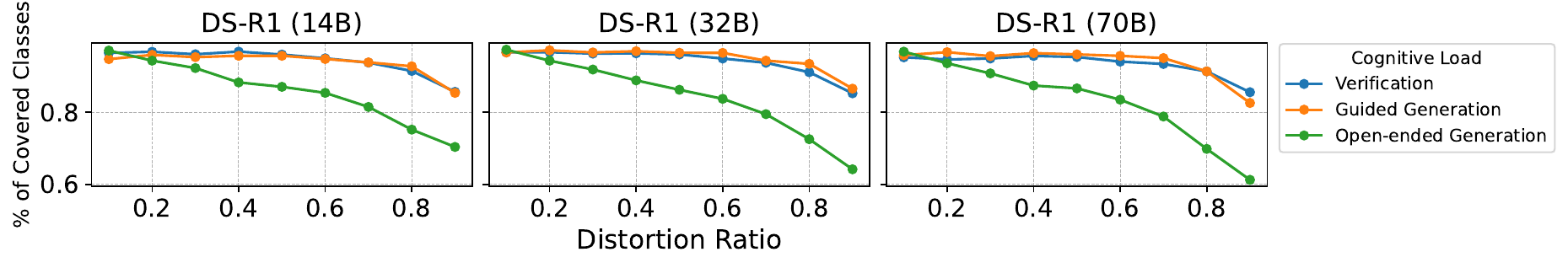}
         \caption{{Open-weight models}}
         \label{fig:covered_classes_vs_dr_coupling_oweight}
     \end{subfigure}
     
     \vspace{1em} 

     \begin{subfigure}[b]{.7\textwidth}
         \centering
         \includegraphics[width=\textwidth]{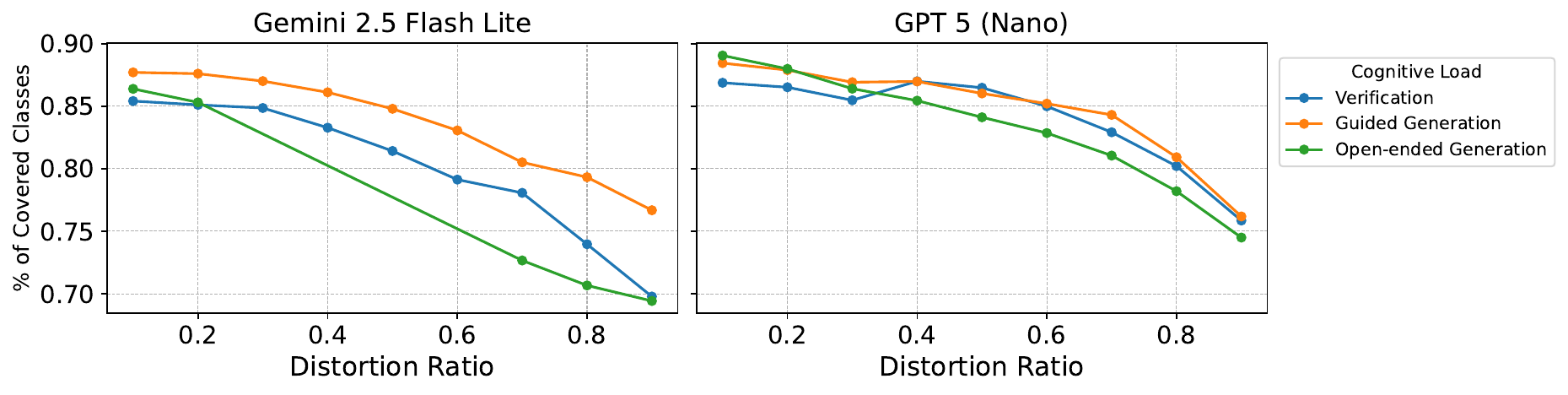}
         \caption{{Proprietary models}}
         \label{fig:covered_classes_vs_dr_coupling_prop}
     \end{subfigure}
     \caption{Relationship between the distortion ratio and the percentage of analyzed classes in the reasoning traces in the coupling experiment. 
     }
     \label{fig:covered_classes_vs_dr_coupling}
\end{figure}

To investigate this discrepancy, we hypothesized that without explicit guidance, the models might not be comprehensively exploring the entire problem space. To test this, we quantified the percentage of classes mentioned in the prompt that were analyzed in the CoT trace as shown in Figure~\ref{fig:covered_classes_vs_dr_coupling}.

The results from this analysis provide a clearer picture. For both \textit{Verification} and \textit{Guided Generation} tasks, the models consistently analyze a high percentage of the provided classes, typically >$90\%$, with only a minor drop-off at the highest distortion ratios. This indicates a methodical and exhaustive approach when the task is well-defined or seeded.

In contrast, for the \textit{Open-ended Generation} task, the fraction of covered classes drops sharply as the distortion ratio increases. For all three models, the coverage begins high but degrades significantly, with the $14$B model dropping to $\sim$$70\%$ and the $70$B model to $\sim$$61\%$ at a distortion ratio of $0.9$. This decline confirms that the models adopt a form of \textit{cognitive shortcut}. Interestingly, when inspecting $20$ random traces, we found that the model does seem to know how to proceed (naïvely at least), by acknowledging that it has to go through each task one by one and inspect its dependencies. However, 
this approach becomes inadequate as the number of classes scales up.

\subsubsection{Trends during cohesion analysis}
Turning to cohesion, the analysis of CoT traces reveals a distinct reasoning pattern that contrasts with the findings for coupling. In this case, the length of the reasoning traces aligns with the expected cognitive load of the task formulations.

\begin{figure}[!t]
     \centering
     \begin{subfigure}[b]{\textwidth} 
         \centering
         \includegraphics[width=\textwidth]{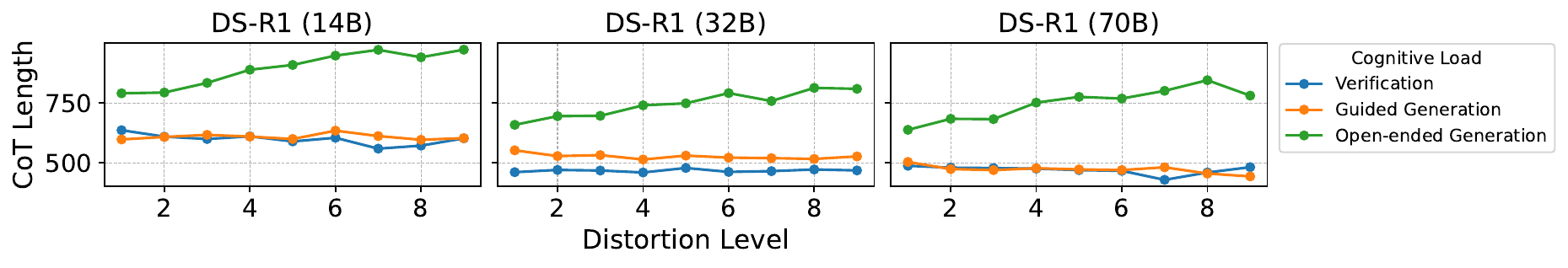}
         \caption{{Open-weight models}}
         \label{fig:cot_len_vs_dr_cohesion_oweight}
     \end{subfigure}
     
     \vspace{1em}

     \begin{subfigure}[b]{.7\textwidth}
         \centering
         \includegraphics[width=\textwidth]{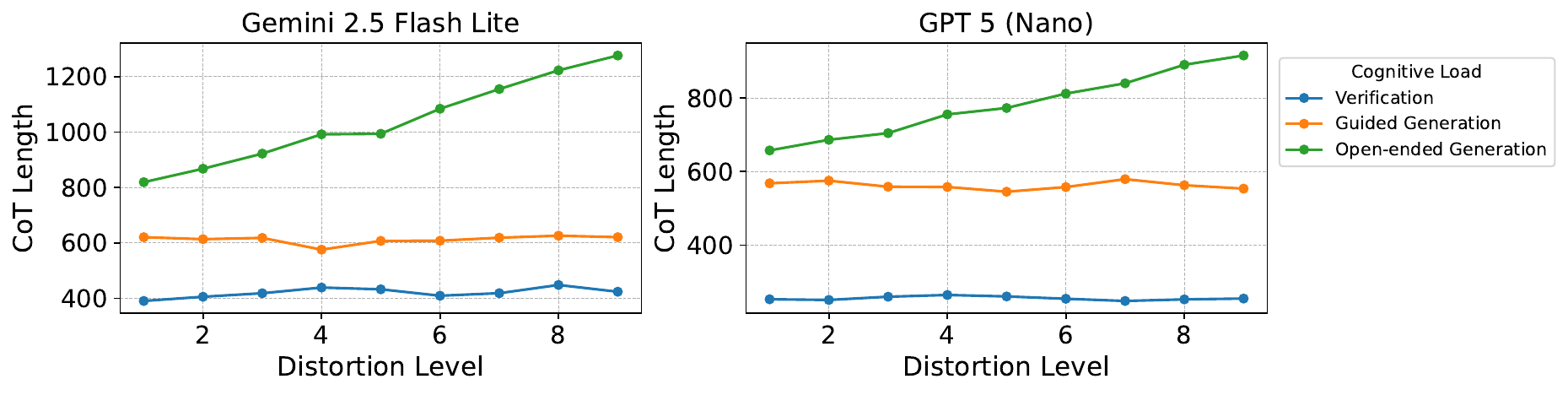}
         \caption{{Proprietary models}}
         \label{fig:cot_len_vs_dr_cohesion_propt}
     \end{subfigure}
     \caption{Relationship between the distortion levels and the length of reasoning traces across all models on the cohesion task.}
     \label{fig:cot_len_vs_dr_cohesion}
\end{figure}

As depicted in Figure~\ref{fig:cot_len_vs_dr_cohesion}, the \textit{Open-ended Generation} task consistently required the most reasoning, producing the longest traces. The \textit{Verification} and \textit{Guided Generation} tasks yielded shorter traces that were not only comparable to each other but also remained stable across all distortion levels. This stability suggests that when the task is highly structured, the models adopt a fixed, procedural approach to analyzing the class, regardless of its internal complexity.

\begin{figure}[H]
    \includegraphics[width=.4\textwidth]{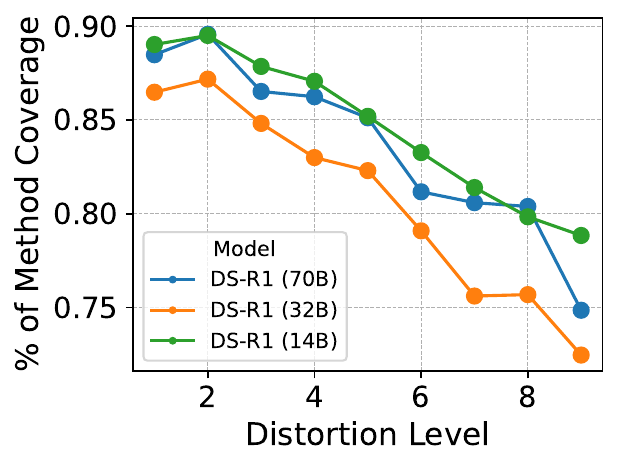}
    \caption{Relationship between method coverage and distortion level in Open-ended Generation {for open-weight models}.}
    \label{fig:cohesion_method_coverage}
\end{figure}

The most telling trend is observed in the open-ended task. Here, the CoT length exhibits a positive correlation with the distortion level; as more unrelated responsibilities are injected into the class, the models generate progressively longer reasoning traces. This indicates that, unlike in the coupling scenario, the models actively attempt to grapple with the increasing difficulty of partitioning the methods. 
To understand the effectiveness of this increased effort, we examined the percentage of methods within the class that were mentioned in the CoT traces for the open-ended task. As shown in Figure~\ref{fig:cohesion_method_coverage}, while the models maintain a relatively high coverage of methods compared to the open-ended coupling task, this coverage still systematically decreases as the distortion level rises. For instance, the $70$B model's coverage drops from $\sim$$88\%$ at the lowest distortion level to $\sim$$75\%$ at the highest.

\subsubsection{Trends observed across coupling and cohesion}

\begin{figure}[H]
    \includegraphics[width=.35\textwidth]{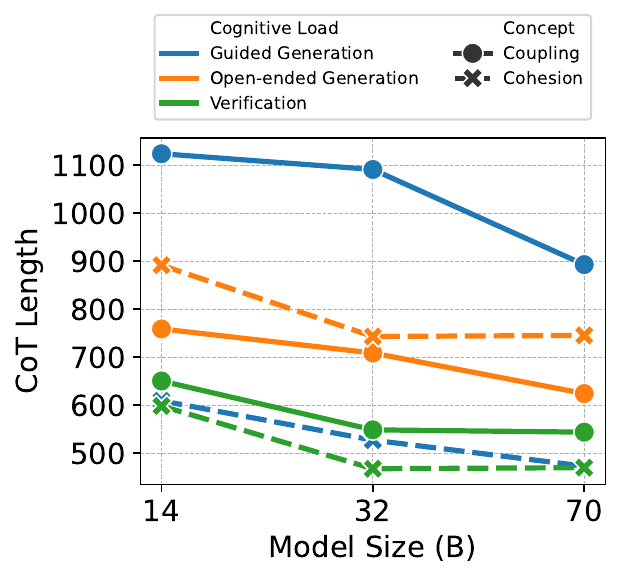}
    \caption{{Open-weight} model size vs. reasoning length. Smaller models tend to generate longer CoT traces.}
    \label{fig:model_size_cot_size}
\end{figure}

A final trend observed across both the coupling and cohesion tasks is the inverse relationship between model size and CoT length. The results, illustrated in Figure~\ref{fig:model_size_cot_size}, consistently show that the larger models, $32$B and $70$B, generate more concise reasoning traces than the smaller $14$B model. This pattern holds true across all three task formulations for coupling and is generally evident for cohesion as well.
This conciseness does not correlate with lower performance; on the contrary, as mentioned in RQs 1-3, the larger models are typically more accurate. This suggests that the reduction in trace length is a signal of greater reasoning efficiency. As model scale increases, their internal representations of software design principles become more sophisticated. Consequently, they can reach conclusions more directly, internalizing the foundational, step-by-step logic that smaller models must articulate explicitly.

\begin{boxH}
\textbf{Summary}: The analysis reveals that though the \llm{}s' problem-solving strategies and failure modes differ between cohesion and coupling,a consistent trend shows that larger models are more efficient, reaching conclusions with shorter, more direct reasoning traces.
\end{boxH}
\section{Discussion}
{In this section, we provide a set of discussion points in light of the findings of the study. First, we touch upon the limits of scalable reasoning that was observed in the open-weight models. Second, we present the implications of our findings on researchers, practitioners, and tool creators. We end the section with a clarification on some anthropomorphic terminology used in our study to avoid any pitfalls.}

\subsection{Limits of Scalable Reasoning}

Our analysis of reasoning traces for the coupling tasks revealed a counterintuitive finding: {open-weight} models produced shorter, less detailed responses for the most complex, open-ended scenarios. {The most consistent explanation with our evidence is} a form of \textit{cognitive shortcutting}, where the models abandon a comprehensive analysis of all provided classes.

\begin{figure}[H]
    \includegraphics[width=.5\textwidth]{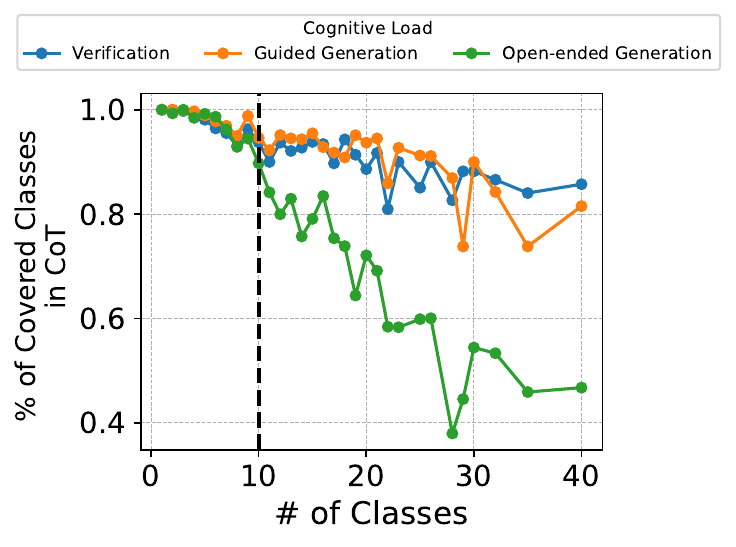}
    \caption{Analyzed class percentage vs. total class count by cognitive level using DeepSeek R1 ($32$B)}    \label{fig:class_coverage_number_of_classes}
\end{figure}

We hypothesize that this failure is a direct consequence of the combinatorial explosion inherent in the task. Identifying all coupled pairs among $N$ classes requires, at a minimum, considering $\binom{N}{2}$ potential relationships. The models appear to default to a naive, brute-force-like strategy that does not scale. As the number of classes increases, this approach becomes intractable, causing the reasoning process to break down. The data from our experiment supports this, as illustrated in Figure~\ref{fig:class_coverage_number_of_classes}: in the constrained \textit{Verification} and \textit{Guided} tasks, class coverage in the reasoning traces remains high, typically above $85-90\%$. In contrast, for the \textit{Open-ended} task, coverage drops sharply as the number of classes grows, falling below $50\%$. 

Despite their large size, these models do not appear to be leveraging efficient heuristics for a problem that, due to our controlled data generation, had a simple and discernible pattern. This reliance on an unscalable, brute-force strategy instead of strategic pattern deduction reveals a bottleneck in their problem-solving abilities. Future work can investigate this phenomenon, as it may be a symptom of a deeper limitation within the models' architecture or training, raising questions about their applicability to any task requiring scalable combinatorial reasoning.

\subsection{{Implications}}

\subsubsection{{Researchers}}

{Cognitive load can be considered as an evaluation axis in \llm{} benchmarks studies. In all three task formulations, the cues for the design issue remain present in the code presented to the model: the dependency injection reversal, the disconnected method clusters. The change is in how much help the prompt provides the model. \textit{Verification} names the pair under question, \textit{Guided Generation} provides the initial code elements the model must consider, and \textit{Open-ended Generation} provides nothing. The drop from \textit{Verification} to \textit{Open-ended} therefore, measures whether the model can locate the relevant structure on its own. A single aggregate score cannot capture this degradation property. Hence, it would be an interesting future work to investigate whether similar observations can be made when models are evaluated on other software engineering tasks under similar experimental settings.}

{Second, our work was confined to recognition, which represents one step, but not the entire process of many activities such as refactoring. A natural extension is to instruct the model to refactor the identified element under the same scaffolding and distortion conditions, and to check whether the resulting code recovers the pre-transformation structure. This would extend the investigation from \textit{can the model recognize the flaw} to \textit{can the model fix it without introducing a new one}.}

{Third, the difference in how distortion affects coupling and cohesion is itself worth studying. Our results showed how coupling performance degrades at nearly every cognitive level under distortion, while cohesion degrades only on \textit{Open-ended Generation}. A plausible explanation is that coupling reasoning relies on a global view of the codebase, whereas cohesion reasoning relies on local signals, such as field access and direct method calls. This degradation caused by the requirement of global reasoning was also observed in similar studies~\cite{Ding2023} where models tend to perform worse in code completion scenarios that require reasoning across multiple files.}

{Finally, generating controlled benchmarks for \llm{}s helps investigate subjective software design properties realistically. The inputs to our transformation are real, mature open-source Java repositories. Each transformation applies the \textit{inverse} of a real-world refactoring: replacing dependency injection with direct instantiation to introduce tighter coupling, or mixing methods from unrelated classes to break cohesion. Because the surrounding code remains exactly as written by the original developers, the resulting evaluation set retains real-world complexity. This procedure can easily extend to other subjective properties such as readability or modularity, offering an alternative to mining real-world labels, which is often noisy and difficult to scale given the need for manual validation. Furthermore, our controlled approach mitigates the risk of data leakage~\cite{Lopez2025}. Evaluating models on directly mined code snippets threatens the reliability of the results, as the models may have encountered those exact snippets during training. This risk is exacerbated when evaluating closed-source models whose exact data provenance remains unknown.}

\subsubsection{{Implications for Tool Creators and Practitioners}}

{The first insight is that bounded questions produce reliable answers as opposed to open-ended explorations. The human--\llm{} gap is small when the prompt names the pair under question or provides a seed, and grows when the model has to find the relevant structure on its own. Tools should let the user point at a class or method and have the model expand from there, rather than scan an entire project for design issues unaided. Practitioners should phrase prompts the same way: ask whether a specific pair is tightly coupled, or which methods relate to a given seed method, rather than asking for every design problem in a module.}

{Second, trimming the prompt to what the question is actually about is important. Coupling performance drops as distractor classes are added at every cognitive level, while cohesion stays stable on \textit{Verification} and \textit{Guided Generation}. Tools should pre-filter the input for coupling-related features using static analysis or a dependency graph. For cohesion-related features, the scope is already local, and tools can include more code in the prompt without losing accuracy. Practitioners should pass only the classes that are directly involved in the question.}

{Third, tools that present a single open-ended pass as the final answer should split the problem into smaller chunks or iterate over the candidates the model returns. This is due to the shortcutting issue that some models exhibit as stated in the results of RQ$5$. Practitioners should re-prompt for classes that were not addressed in the first pass and verify each returned candidate before acting on it.}

{Furthermore, no single model is best on both concepts. DeepSeek-R1 ($70$B) leads on coupling. GPT-$5$ Nano and Gemini-$2.5$ Flash Lite lead on cohesion under \textit{Open-ended Generation}. GPT-$5$ Nano leads on cohesion \textit{Verification} but is the weakest model on coupling \textit{Open-ended Generation}. Tools that address problems related to such design concepts should route different tasks to different models.}

{Finally, the output of the model should be treated as a suggestion and the final decision should be left to the human developer. The human--\llm{} gap is largest in the hardest experimental setting, which is also the closest to a real refactoring decision. Whether the integration is a code-review bot, an IDE plugin, or a developer pasting code into a chat window, the model's role is to surface candidates that a human evaluates. It is not a substitute for that evaluation.}

\subsection{Anthropomorphization and Large Language Models}

In this study, we employed terminology traditionally rooted in human psychology, 
such as ``\textit{cognitive load}'' and a framework inspired by Bloom's 
Taxonomy~\cite{Bloom1964}. We clarify that our use of these terms is 
functional and metaphorical, rather than an attribution of human-like intent to \llm{}s.

We use ``\textit{cognitive load}''  as a proxy for the information processing burden placed on the model. ``\textit{High cognitive load}'' scenarios (\eg{} Open-ended Generation) correspond to tasks with a larger search space, requiring the model to sustain attention across context without the scaffolding of specific retrieval cues. Conversely, ``\textit{Verification}'' acts as an easier task where the search space is collapsed to a binary classification.

Furthermore, our adaptation of Bloom's Taxonomy serves as a structural framework to graduate the \textit{level of guidance} provided in the prompt. It allows us to distinguish between a model's ability to recognize a pattern when explicitly pointed out (\textit{Verification}) versus its ability to reconstruct that pattern from a noisy input (\textit{Creation}).

Moreover, we observed that models produced shorter reasoning traces in complex coupling tasks. We term this ``\textit{cognitive shortcutting}'' not to imply laziness, but to describe a failure mode where the model's generation converges prematurely when it is required to process combinatorial relationships.

Finally, we clarify our use of the \textit{intern} versus \textit{Senior Engineer} analogy. This comparison is strictly operational regarding autonomy and ambiguity tolerance, rather than a reflection on the competence of early-career professionals. In our work, an \textit{intern} is capable of high efficiency when executing well-bounded tasks under supervision, whereas a \textit{Senior Engineer} is capable of navigating noisy, open-ended {design analysis tasks} independently. Our results suggest that while \llm{}s exhibit the execution speed of the former, they lack the self-correction and contextual span required of the latter.

{We further note that the analogy is restricted to the kinds of tasks evaluated in this study, namely the recognition, retrieval, and partitioning of design elements. We do not claim that our findings extend to end-to-end refactoring competence, the weighing of design trade-offs, or architectural decision-making, which are out of scope. However, refactoring and software design improvement are multi-step processes whose first step is the identification of design issues, with later stages depending on the soundness of that identification~\cite{Leppanen2015, Haendler2018}. Fragility at such a stage, therefore, may affect the reliability of any downstream activity built on top of it.}
\section{Limitations and Threats to Validity}
This section discusses the potential limitations of our study and the steps taken to mitigate them.

\textbf{Construct Validity}: This threat concerns whether our methods for creating low-cohesion and high-coupling examples accurately represent these abstract software design concepts. To ensure our constructs were well-grounded, we based our data generation on established principles: (1) for cohesion, we started with real-world Java classes confirmed to be highly cohesive (YALCOM=0 and method count greater than 2) before systematically injecting noise, (2) for coupling, we implemented transformations that directly loosen the coupling by replacing dependency injection with direct instantiation. As for the evaluation metrics, the F1 score was chosen for classification and retrieval-oriented tasks as it provides a balanced measure of precision and recall. For the open-ended cohesion task, which is a partitioning problem, the Adjusted Rand Index (ARI) provides a more robust evaluation by measuring the similarity of method clusters while correcting for chance agreements.

\textbf{Internal Validity} addresses whether the observed outcomes are truly caused by our independent variables (distortion ratio, level of help) or by other confounding factors, such as the inherent complexity, naming conventions, or token length of the code snippets. We minimized the influence of confounding variables through our highly controlled, programmatic data generation pipeline. To reduce the potential effect of code length, we employed stratified sampling across the token length of our generated examples. This ensured that the distribution of code snippets was balanced across all experimental conditions. This step allowed us to isolate the impact of our primary variables (distortion and guidance) from the confounding effect of code length. In addition, our experiments were conducted with the temperature parameter set to 0 to ensure reproducibility and eliminate randomness as a confounding variable.
\added{A further threat to internal validity arises from our inclusion of closed-source models. Unlike the open-weight models, which are immutable upon release, proprietary models are subject to often undocumented updates, and eventual deprecation~\cite{angermeir2025reflections, Abdurahman2024, Spirling2023, KapoorNarayanan2023OpenAI}. These updates may alter model behavior and could undermine the long-term replicability of our findings. We mitigated such a threat by carrying out our experiments within a narrow time window against a stable (not preview) version where the provider's API exposed such versioning. Nevertheless, we acknowledge that exact re-execution of the closed-source portion of our experiments may not be possible once the corresponding endpoints are retired.}
{Regarding the construction process of the cohesion ground truth, the criterion in Section $3.2.4$ guarantees that source classes are structurally separable. However, it does not guarantee that they are \textit{always} conceptually unrelated since we rely on the structural definition of cohesion found in the literature~\cite{Sharma2020, HendersonSellers1996CouplingAC, Chidamber1994, Hitz1996, Li1993}. We have done so out of a practical need to generate instances at scale. We leave a systematic characterization of such edge cases to future work.}

\textbf{Ecological Validity} examines whether the findings of a study can be generalized to naturalistic situations. In our work, we aimed for high internal validity through our controlled experimental design, {and maximizing both internal and ecological validity is impossible~\cite{StolFitzgerald2018, McGrath1981}}. However, to reduce threats to ecological validity, we mined our samples from real-world repositories that were written by actual developers to satisfy realistic use cases. In addition, our noise injection process further emphasizes the realism of our evaluation. This is in contrast to similar published work~\cite{Du2024} that benchmarks \llm{}s in a contextualized, isolated environment, where isolated classes are evaluated for a simple autocompletion task with problems framed from an undergraduate course. Such a setup does not capture the reality of repository-level development with thousands of dependencies and noise in the form of unrelated code w.r.t a class. Our methodology addresses this gap by operationalizing noise as the presence of irrelevant context. We aimed to simulate the burden of real-world architectural reasoning that developers encounter when navigating codebases. The cohesion noise is also realistic as it simulates how real-world classes often accumulate unrelated responsibilities over time, as demonstrated by~\cite{Chatzigeorgiou2014}. Injecting methods from distinct, unrelated source classes simulates such drift. 
{A related threat concerns the abstract representation of the inputs presented to the models. For the coupling experiments, classes are skeletonized: method bodies that do not involve dependency instantiation are removed, and only signatures are retained. This abstraction is necessary because objective ground-truth labels for structural coupling require us to guarantee the absence of latent coupling between any two classes, which intact method bodies cannot ensure. We acknowledge that this abstraction is different from how developers typically encounter code in practice and may bias model reasoning by removing semantic cues. Two considerations mitigate this concern. First, the cohesion experiments retain full method bodies, comments, and original method names, and exhibit the same degradation patterns under reduced guidance and increased distortion. The trends we report are therefore unlikely to be artifacts of skeletonization. Second, skeletonization yields a more controlled and focused setting than non-abstract code, with smaller inputs and no distracting in-method content. If models already fail in this controlled setting, they are likely to perform worse on the non-abstract representation, so the fragility we observe is plausibly a \textit{upper bound} on what would be observed with richer inputs.}

\textbf{External Validity} relates to the generalizability of our findings. The results, derived from Java code, may not apply to other programming languages (\eg{} Python, C++), different programming paradigms (\eg{} functional). Yet, we sought to improve generalizability by selecting Java, a mainstream object-oriented language, and by evaluating a {set} of models at three distinct scales ($14$B, $32$B, and $70$B), with heterogeneous foundational models (Qwen-$2.5$ and Llama $3.3$), {including proprietary ones}. Observing consistent trends across these model sizes provides stronger evidence than a single-model study. Furthermore, we expect RQ$1$–RQ$3$ trends to hold across model families given the neural scaling laws~\cite{Bahri2024, Hoffmann2022} that establish a relationship between model size and performance. In addition, Wu~\etal{}~\cite{Wu2025when} have shown that such models gravitate toward shorter CoTs as they become more capable, as was observed in our results in RQ5. {Finally, in this study, we only considered \textit{structural} coupling and \textit{functional} cohesion. We did so to keep the scope of the study within reasonable bounds and to take advantage of the convenience of violating such principles programmatically. We leave the remaining types for future work.}
\section{Conclusions}
In this work, we sought to investigate the depth and robustness of \llm{}'s understanding of cohesion and coupling. Through a controlled experiment using generated code with varying levels of noise and contextual guidance, we mapped the boundaries of the model's {performance on design analysis tasks, namely the recognition, retrieval, and partitioning of design elements}. Findings reveal that these models possess a solid foundational knowledge of both principles but their practical ability to apply these principles is brittle. They are more adept at identifying inter-module coupling but are highly vulnerable to contextual noise, whereas their grasp of intra-module cohesion is more resilient but falters when tasks require autonomous discovery. Ultimately, our results indicate that current models perform well under guidance but lack the robust, scalable reasoning to operate in unconstrained {design analysis} scenarios. Future work could investigate the reasons, from an architectural or data viewpoint, for why they perform as such. In addition, applying our methodology that incorporates different guidance and distortion levels in other SE tasks would be of interest.

\begin{acks}
This work was partially funded by the Natural Sciences and Engineering Research Council of Canada (NSERC) through grant NSERC Discovery RGPIN-$2023$-$04903$, and partially supported by the Wallenberg AI, Autonomous Systems and Software Program (WASP) funded by the Knut and Alice Wallenberg Foundation, Software Center Project \#$61$ and the Vinnova competence center on Continuous Digitalization (CoDiG). Jos\'e Antonio Hern\'andez L\'opez's work was supported by the Project PID2022-140109NB-I00 funded by MICIU/AEI/10.13039/501100011033 and by ERDF, EU, and Project PID2025-169717NB-I00 funded by MICIU/AEI/10.13039/501100011033.

\end{acks}
\bibliographystyle{ACM-Reference-Format}
\bibliography{references}

\end{document}